\documentclass[graphics,floatfix, footinbib,tightenlines,nobibnotes, aps, prb, twocolumn]{revtex4-1}
\usepackage{amsmath,amssymb}
\usepackage{graphicx}
\usepackage{dcolumn}
\usepackage{bm}
\usepackage{braket}
\usepackage{subcaption}
\usepackage{verbatim}
\usepackage{float}
\usepackage{bbold}
\usepackage{color}
\usepackage{xcolor}
\usepackage{relsize}
\usepackage{amsthm}
\usepackage{enumerate}
\usepackage{soul,xcolor}
\usepackage[T1]{fontenc}
\usepackage[colorlinks=true ,urlcolor=blue,urlbordercolor={0 1 1}]{hyperref}

\newcommand{\beq}{\begin{eqnarray}}
\newcommand{\eeq}{\end{eqnarray}}

\newcommand{\bsp}{\begin{split}}
\newcommand{\esp}{\end{split}}

\newcommand{\be}{\begin{equation}}
\newcommand{\ee}{\end{equation}}

\begin{document}

\setstcolor{red}

\title{Spin Liquids and Pseudogap Metals in $SU(4)$ Hubbard Model in Moir\'e Superlattice}
\author{Ya-Hui Zhang and Dan Mao}
\affiliation{Department of Physics, Massachusetts Institute of Technology, Cambridge, MA, USA
}

\date{\today}

\begin{abstract}
 Motivated by the  realization of spin-valley Hubbard on triangular moir\'e superlattice in ABC trilayer graphene aligned with hexagon boron nitride (hBN) and possibly also in twisted transition metal dichalcogenide homobilayers, we study possible Mott insulating phases and pseudogap metals based on symmetry constraint and parton mean field theories. First we show that Luttinger constraint allows two distinct symmetric and featureless Fermi liquids when there is an inter-valley Hund's term breaking $SU(4)$ spin rotation. Especially, there exists a symmetric and featureless "pseudogap metal" with small Fermi surfaces.  Then we suggest to search for such an unconventional metallic state by doping the Mott insulator at $\nu_T=2$.  For this purpose, we study the $\nu_T=2$ Mott insulator using $SO(6)$  Schwinger boson or  Schwinger fermion parton. At the $SU(4)$ symmetric point, we find two symmetric $Z_2$ spin liquids. With a large anti inter-valley Hund's term, a  featureless Mott insulator  is natural. Next we show that doping the featureless Mott insulator or a $Z_2$ spin liquid can lead to featureless or orthogonal "pseudogap metal" with small Fermi surfaces proportional to the doping. Besides, we also provide one scenario for the evolution from "pseudogap metal" to the conventional Fermi liquid through an intermediate exotic "deconfined metal" phase.  Last, we give brief comments on the possibility of $U(1)$ spinon fermi surface state or $Z_4$ spin liquid at $\nu_T=1$.
\end{abstract}

\pacs{Valid PACS appear here}
,
\maketitle

\section{Introduction}
Recently moir\'e superlattices from Van der Waals heterostructures emerge to be a wonderful platform  to study strongly correlated physics. These include correlated insulator\cite{cao2018correlated}, superconductivity\cite{cao2018unconventional,yankowitz2019tuning,lu2019superconductors} and anomalous Hall effect\cite{Aaron2019Emergent} in twisted bilayer graphene, spin-polarized correlated insulators\cite{Shen2019Observation, Liu2019Spin,Cao2019Electric}  and superconductivity\cite{Shen2019Observation, Liu2019Spin} in twisted bilayer-bilayer graphene. In addition, ABC trilayer graphene aligned with a hexagonal boron nitride (TLG-hBN) has been demonstrated to host gate tunable correlated insulator\cite{chen2018gate}, superconductor\cite{Wang2019Signatures} and Chern insulator\cite{chen2019tunable}.  

Theoretically it has been shown that both bandwidth and band topology can be tuned by the displacement field in the TLG-hBN system\cite{zhang2019nearly,chittari2019gate}. For one sign of displacement field, the valence bands from the two valleys have non-zero and opposite Chern numbers.  Similar narrow Chern bands have also been predicted in twisted bilayer graphene aligned with h-BN\cite{zhang2019twisted,bultinck2019anomalous} and in twisted bilayer-bilayer graphene\cite{zhang2019nearly,chebrolu2019flatbands,choi2019intrinsic,lee2019theory,koshino2019band,liu2019quantum}. These systems therefore may realize interesting "quantum Hall" physics. Indeed, anomalous Hall effect\cite{Aaron2019Emergent} and Chern insulator with $\sigma_{xy}=2\frac{e^2}{h}$\cite{chen2019tunable} have already been reported.  In contrast, for the other sign of displacement field in TLG-hBN, the valence band is trivial and there is no obstruction to build lattice models from constructing Wannier orbitals\cite{po2018origin}. Such a spin-valley Hubbard model on triangular lattice is derived in Ref.~\onlinecite{zhang2019bridging}. $\frac{t}{U}$ in this Hubbard model can be tuned by the magnitude of the displacement field. Therefore the trivial side of TLG-hBN offers an amazing platform to study Hubbard model physics\cite{xu2018topological,zhang2019bridging,zhu2018antiferro,jian2018moire,schrade2019spin}, which may be similar to  that of the cuprates. The observation of a superconductor in the trivial side\cite{Wang2019Signatures} is encouraging. In cuprates, the pseudogap metal and the strange metal remain as unsolved mystery in addition to the high Tc superconductor. Then a natural question is: can TLG-hBN also host similar "pseudogap metal" and "strange metal" phases?  In this article we try to give a postive answer to this question by explicitly constructing several simple "pseudogap metal" ansatz in the spin-valley Hubbard model.  A $SU(4)$ Hubbard model on triangular lattice may also be realized in twisted transition metal dichalcogenide(TMD) homobilayer\cite{wu2019topological}. Therefore our discussions can also be relevant to future experiments in twisted TMD bilayers.

In cuprates, a sensible theoretical scenario is that the strange metal is associated with a quantum critical point between the pseudogap metal and the conventional Fermi liquid. However, the critical point, even if exists, is covered by the superconducting phase. When the superconductor is suppressed by strong magnetic feild,  in the under-doped region experiments  observe signatures of small Fermi surfaces through quantum oscillation\cite{sebastian2010metal} and Hall measurement\cite{badoux2016change}. The area of the small Fermi pocket inferred from the experiment is proportional to the doped additional holes instead of all of the electrons. It is still under debate whether this high field "pseudogap metal" is from some density wave orders\cite{harrison2011protected,sebastian2012towards} or is from a symmetric metal like FL* phase\cite{senthil2003fractionalized,sachdev2016novel}. As a matter of principle, density wave order parameter is not necessary to gap out Fermi surface and there should exist symmetry "pseudogap metal" with small Fermi surfaces  once fractionalization is allowed. However, we do not know any simple model so far to realize these symmetric metals with small Fermi surfaces.   In this paper we will show that  spin-valley Hubbard model is very promising in this direction. More specifically, we show that at filling $\nu_T=2-x$, there are naturally symmetric "pseudogap metals" with Hall number $\eta_H=-x$. Depending on the value of inter-valley Hund's coupling, the "pseudogap metal" is either a featureless Fermi liquid or an orthogonal metal\cite{nandkishore2012orthogonal}.

We can understand the existence of "pseudogap metals" from two different perspectives. First, with an inter-valley Hund's term, the $U(4)$ symmetry is broken down to $(U(1)\times U(1)_{valley}\times SU(2)_{spin})/Z_2$. Then Lieb-Schultz-Mattis (LSM) constraint allows two distinct symmetric and featurless Fermi liquids with Fermi surface areas $A_{FS}=\frac{2-x}{4}$ or $A_{FS}=-\frac{x}{4}$. In the second perspective, symmetric pseudogap metals can be constructed from doping symmetric Mott insulators.  Therefore we turn to study the possible symmetric Mott insulators at $\nu_T=2$. Depending on the value of inter-valley Hund's coupling $J_H$, we find  a featureless insulator and two symmetric $Z_2$ spin liquids using $SO(6)$ Schwinger boson or Schwinger fermion parton construction.  Then within a $SO(6)$ slave boson parton theory, doping the Mott insulator leads to a featureless Fermi liquid or an orthogonal metal. Both of them have small Fermi surfaces with area equal to $\frac{x}{4}$, resembling experimental results of under-doped cuprates under strong magnetic field.  Compared to phenomenology in cuprates, the ansatz we propose here is much simpler: it is a ground state at zero magnetic field without breaking any symmetry.  The simplicity of the proposed pseudogap metal may make it much easier to study its evolution towards the large fermi surface Fermi liquid and possible "strange metal" phase sandwiched in the intermediate region.  We suggest one possible route through an intermediate "deconfined metal" with an internal $U(1)$ gauge field. It remains a question whether a direct transition is possible or the property of the intermediate phase(or critical region) can mimic that of the strange metal in the cuprates. 

In this paper we focus on the limit that the anisotropic term breaking SU(4) spin rotation symmetry is small compared to the Heisenberg coupling. If the inter-valley Hund's coupling is large, then the $\nu_T=2$ Mott insulator has $120^\circ$ Neel order \cite{thomson2018triangular} formed by spin one moment. Physics from doping such a spin one Neel order  may also be interesting, but is beyond the scope of the current paper.

Although most part of the paper is focused on filling close to $\nu_T=2$. We also give a brief discussion on the Mott insulator at $\nu_T=1$. At $\nu_T=1$, we only find two symmetric spin liquids: a $U(1)$ spinon Fermi surface state or a $Z_4$ spin liquid.  A plaquette order may be a strong competing candidate. With only nearest neighbor coupling, magnetic order may be suppressed by strong quantum fluctuations. Therefore we expect the $\nu_T=1$ Mott insulator to preserve the  approximation $SU(4)$ spin rotation symmetry.  Then  a charge $4e$ superconductor may emerge from doping such a $SU(4)$ symmetric Mott insulator. It is interesting to study the possibility that the observed superconductor in TLG-hBN\cite{Wang2019Signatures} is a charge $4e$ paired state.

\section{Hamiltonian and Symmetry},

A lattice model for TLG-hBN has been derived in Ref.~\onlinecite{zhang2019bridging}. To leading order it is a spin-valley model on triangular lattice:
\begin{align}
	H&=-t \sum_{a}\sum_{\langle ij \rangle}e^{i \varphi^a_{ij}}c^\dagger_{a;i}c_{a;j}+h.c.\notag\\
	& + \frac{U}{2}n_i^2+J_H \sum_i \mathbf{S_{+;i}}\cdot \mathbf{S_{-;i}}
\end{align}
where $a=+,-$ is the valley index. We have suppressed the spin index. $U$ is the Hubbard interaction and $J_H$ is an on-site inter-valley spin-spin coupling.  $\varphi^+_{ij}=-\varphi^{-}_{ij}$ provides the valley contrasting staggered flux pattern.

At $\varphi^a_{ij}=0$ and $J_H=0$ limit, we have $U(4)$ symmetry. Adding the valley-contrasting flux breaks the symmetry down to $U(2)_+ \times U(2)_-$, which is further broken down to  $SU(2)_s\times U(1)_c \times U(1)_v/Z_2$ by the inter-valley spin-spin coupling, where $Z_2$ stands for the common element of $SU(2)_s$, $U(1)_c$ and $U(1)_v$.  The Coulomb interaction indicates that $J_H<0$. However, electron phonon coupling from phonon at $K$ and $K'$ can mediate positive $J_H$\cite{dodaro2018phases}. The final sign of $J_H$ is decided by the competition between these  two effects. In this paper we will view $J_H$ as a phenomenological parameter to be fit from the experiment.

Next we discuss the effective low energy model in the $U>>t$ limit with a restricted Hilbert space. $\nu_T\leq 2$ can be mapped to $\nu_T\geq 2$ by a particle-hole transformation and thus we only focus on $\nu_T\leq 2$. 

\subsection{Mott Insulator}

At integer $\nu_T=1,2$, the charge is localized and the low energy is described by an effective spin model. The dimension of the Hilbert space at each site is $4$ and $6$ (6=4 choose 2) respectively for $\nu_T=1$ and $\nu_T=2$. In the $SU(4)$ symmetric limit, we have
\begin{equation}
 	H_S=J \sum_{\langle ij \rangle}\sum_p S^p_i S^p_j
 \end{equation}
 with $J=\frac{t^2}{2U}$. $S^p_i$ with $p=1,2,...,15$ is an spin operator on each site. These 15 spin operators can be organized as $S^{\mu \nu}=\tau^\mu \sigma^\nu$ with $\mu,\nu=0,1,2,3$ expect $\mu=\nu=0$.  Each of them is a fermion bilinear:
\begin{equation}
 S^{\mu\nu}_i=\sum_{a_1,a_2}\sum_{\sigma_1\sigma_2} c^\dagger_{a_1 \sigma_1;i}\tau^\mu_{a_1 a_2} \sigma^\nu_{\sigma_1,\sigma_2}c_{a_2,\sigma_2;i}
 \label{eq:spin_operator_definition}
\end{equation}
  Projecting to the four and six dimensional Hilbert space at each site for $\nu_T=1,2$, $S^{\mu\nu}=\tau^\mu \sigma^\nu$ should be viewed as  $4\times 4$ and $6 \times 6$ matrices for $\nu_T=1$ and $\nu_T=2$. 

 The spin Hamiltonian has $PSU(4) = SU(4)/Z_4$ symmetry.  Here $Z_4$ means the global $U(1)$ transformation $e^{i\frac{2\pi}{4}n}$ with $n$ as an integer.  At $\nu_T=2$, each site is in the $6d$-irrep of $SU(4)$, which transforms like $SO(6)$ fundamental under $SU(4)$ ($SO(6)\cong SU(4)/Z_2$).

For TLG-hBN system, there is a valley contrasting phase in the hopping term\cite{zhang2019bridging}. Two valleys have opposite staggered flux patterns. This valley dependent flux on the hopping term is inherited in the $t^2/U$ expansion and gives an anisotropy term:
\begin{align}
	H'_S&=J \sum_{\langle ij \rangle}(\cos 2\varphi_{ij}-1)(\tau^x_i \tau^x_j+\tau^y_i \tau^y_j)(1+\vec{\sigma}_i\cdot \vec{\sigma}_j) \notag\\
	&+J \sum_{\langle ij \rangle}\sin 2\varphi_{ij}(\tau^x_i \tau^y_j-\tau^y_i \tau^x_j)(1+\vec{\sigma}_i\cdot \vec{\sigma}_j) 
\end{align}
where $\varphi_{ij}$ is the phase in the hopping for the valley $+$ on bond $\langle ij \rangle$ (Correspondingly the valley $-$ has phase $-\varphi_{ij}$ for the hopping on the same bond).

The above symmetry breaks the $PSU(4)$ spin rotation down to  $SO(3)_+\times SO(3)_-\times U(1)_v$. For $\nu_T=2$ the corresponding spin rotation symmetry can be viewed as $U(1)_v\times SO(4)/Z_2 \cong SO(3)_+\times SO(3)_-\times U(1)_v$, where $SO(4)$ acts on the $4$d space formed by three valley singlet, spin triplet and one valley triplet, spin singlet. 

For $\nu_T=2$, there is an additional on-site inter-valley spin-spin coupling:
\begin{equation}
	H''_S=J_H\sum_i \vec{S}^+_i \cdot \vec{S}^-_i
\end{equation}

We define the valley specified spin operator:
\begin{equation}
	\vec{S}^a_i=\frac{1}{2}\sum_{\sigma_1,\sigma_2=\uparrow,\downarrow}c^\dagger_{i;a;\sigma_1} \vec{\sigma}_{\sigma_1 \sigma_2} c_{i;a;\sigma_2}
\end{equation}
for $a=+,-$.

$H''_S$ vanishes for $\nu_T=1$. For $\nu_T=2$, it further breaks the spin rotation symmetry down to $U(1)_{valley}\times SO(3)_{spin}$. 

\subsection{Finite doping: type I and type II $t-J$ models}

\begin{figure}[ht]
\centering
\includegraphics[width=0.45\textwidth]{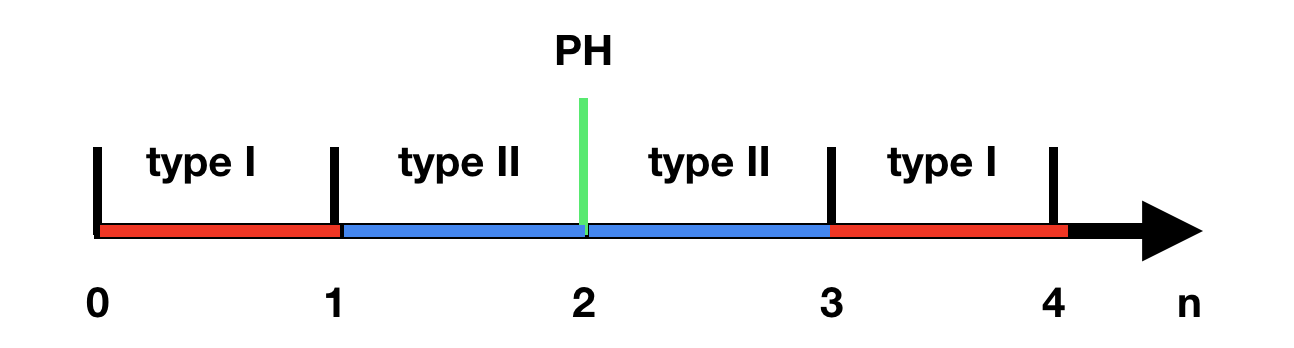}
\caption{Two distinct $t-J$ models in four-flavor spin-valley Hubbard model. $\nu_T<2$ and $\nu_T>2$ are related by particle-hole transformation.  $\nu_T<1$ and $1<\nu_T<2$ realize two essentially different $t-J$ models.}
\label{fig:two_t_J}
\end{figure}

In spin $1/2$ Hubbard model, the physics at finite doping away from the Mott insulator is believed to be governed by a $t-J$ model at the $U>>t$ limit. Here we want to show that for the spin-valley Hubbard model, the region $1<\nu_T<2$ has different physics from the traditional $t-J$ model in the region $0<\nu_T<1$. Therefore the four flavor Hubbard model can actually realize two distinct $t-J$ models, which is illustrated in Fig.~\ref{fig:two_t_J}.

\begin{figure}[ht]
\centering
\includegraphics[width=0.45\textwidth]{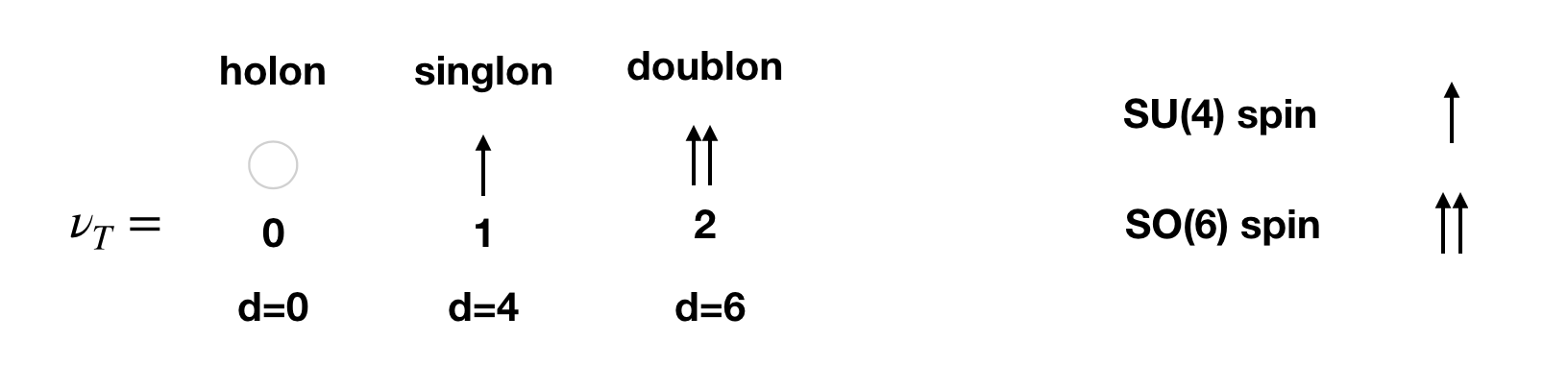}
\caption{Label of several different states according to the number of particle $n_i$ at the site. Holon is an empty site. Singlon and doublon are in the fundamental rep and the $SO(6)$ rep of the $SU(4)$ group respectively. The Hilbert space dimensions of the holon, singlon, doublon states are $1$, $3$ and $6$ respectively.}
\label{fig:holon_singlon_doublon}
\end{figure}

\subsubsection{$\nu_T=1-x$: type I $t-J$ model}
At filling $\nu_T=1-x$ with $0<x<1$, the low energy is described by a similar $t-J$ model as in the spin $1/2$ case:
\begin{equation}
	H=-t \sum_{\langle ij \rangle} P_1 c^\dagger_i c_j P_1+J \sum_{\langle ij \rangle}S^p_i S^p_j+H'_S
	\label{eq:conventional_t_J}
\end{equation}
where $P_1$ is the projection operator which forbids states with $n\geq 2$ on each site.

For $\nu_T=1-x$, the on-site inter-valley spin-spin coupling $J_H$ term vanishes in the leading order of $t/U$  because of the restriction of the Hilbert space.  It enters in the second order of $t/U$ by changing the  spin coupling from $\frac{t^2}{U}$ to $\frac{t^2}{U \pm J_H}$. The change of the spin coupling is $\delta J\sim \frac{J_H}{U} J<<J$ and can be ignored give the estimation that $J_H \sim 0.01 U$\cite{zhang2019bridging}. Therefore there is an approximating $U(2)_+\times U(2)_-$ symmetry at the $U>>t$ limit even if $J_H \neq 0$.

In this type I $t-J$ model, each site is either empty or singly occupied, similar to that of hole-doped cuprate\cite{lee2006doping}.   For convenience, in this paper we call the empty site as holon and the singly occupied site as singlon (see Fig.~\ref{fig:holon_singlon_doublon}).  The Hilbert space dimension of each site is $1+4=5$. The $t-J$ model has unusual property that the singlon density is conserved to be $1-x$ while the holon density is conserved to be $x$. The $t$ term in Eq.~\ref{eq:conventional_t_J} is not a traditional hopping term. Instead, it involves the exchange between a holon at site $i$ with a singlon at site $j$.

\begin{figure}[ht]
\centering
\includegraphics[width=0.45\textwidth]{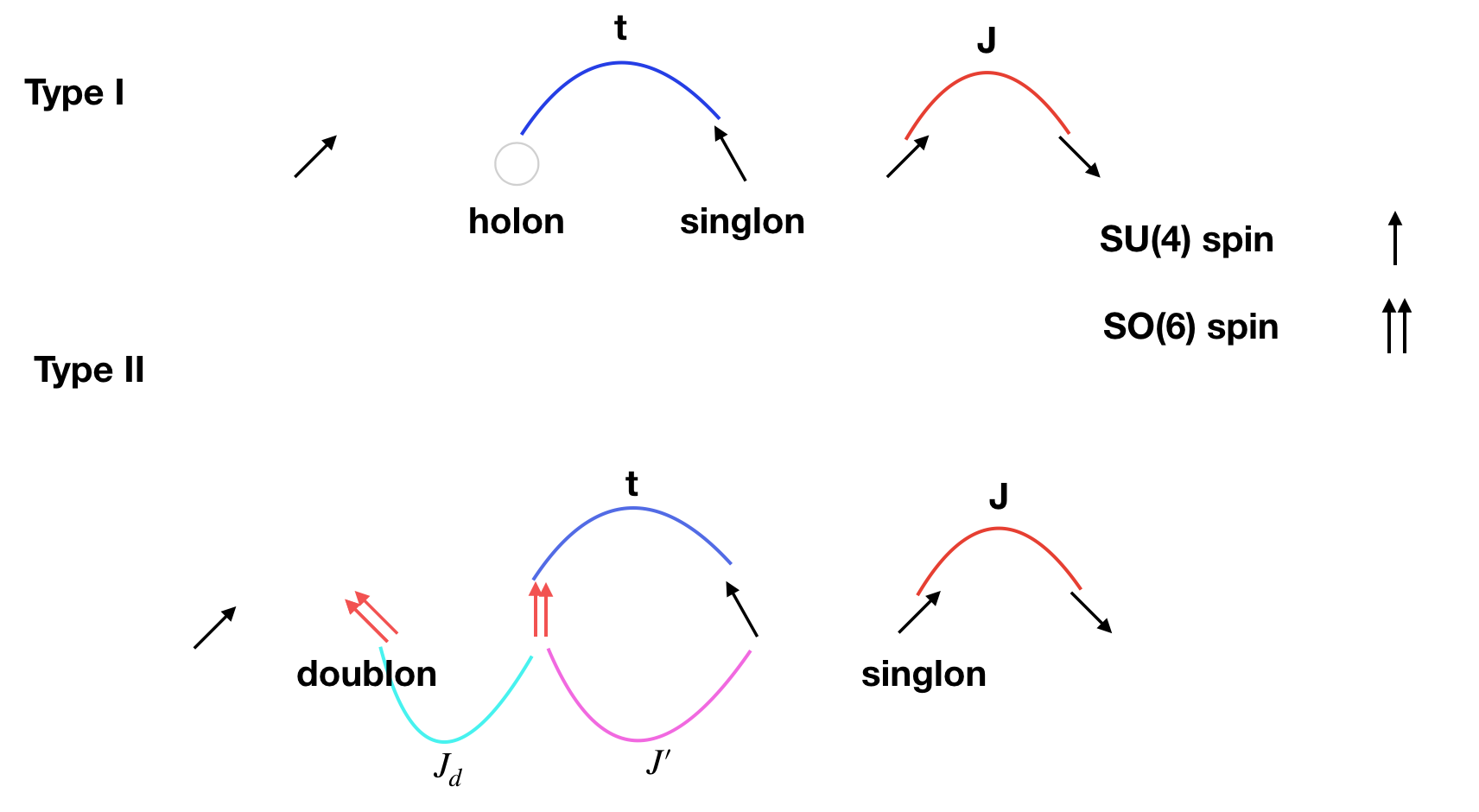}
\caption{Illustration of the type I and type II $t-J$ models. In type II model, there are three different spin-spin couplings.}
\label{fig:type_I_II_t_J}
\end{figure}

\subsubsection{$\nu_T=2-x$: type II $t-J$ model}
At filling $\nu_T=2-x$ with $0<x<1$, we have either a singlon or a doublon state at each site in the $U>>t$ limit. Thus the Hilbert space dimension at each site is $4+6=10$. In addition, the singlon and the doublon carry different representations of $SU(4)$ spin. Hence there are three different spin couplings (see Fig.~\ref{fig:type_I_II_t_J}). We define  $P_s$ and $P_d$ as the projection operators to the singlon and the doublon states respectively. Then it is natural to have spin operators for the singlon and doublon: $S^{\mu\nu}_{i;s}=P_s S^{\mu \nu}_i P_s$ and $S^{\mu\nu}_{i;d}=P_d S^{\mu \nu}_i P_d$.  We have the type II $t-J$ model as

\begin{align}
	H&=-t \sum_{\langle ij \rangle} (P_s+P_d) c^\dagger_i c_j (P_s+P_d)+J \sum_{\langle ij \rangle}S^p_{i;s} S^p_{j;s}\notag\\
	&+\frac{1}{2} J' \sum_{\langle ij \rangle} (S^p_{i;s} S^p_{j;d}+S^p_{i;d} S^p_{j;s})+J_d\sum_{\langle ij \rangle}S^p_{i;d} S^p_{j;d} \notag\\
	&+H'_S+H''_S
	\label{eq:unconventional_t_J}
\end{align}
where  $J'=\frac{1}{2}J$ and $J_d=J$. In the super-exchange process involving a singlon and a doublon nearby, the only process we should keep is to hop the particle from the singlon to the doublon, which costs energy $2U$ instead of $U$. This is how the two factor $1/2$ arise. Because of the $H''_S$ term, generically we only have $U(2)\times U(1)_{valley}$ spin rotation symmetry.

The two $t-J$ models in Eq.~\ref{eq:conventional_t_J} and Eq.~\ref{eq:unconventional_t_J} are apparently quite different. In the type II t-J model, there is no empty site in the Hilbert space. Instead, the singlon and doublon both carry spin. The kinetic term now becomes the exchange of the singlon and doublon.  Recently, a similar type II $t-J$ model has been proposed\cite{zhang2019kondo} to describe the nickelate superconductor\cite{li2019superconductivity}. There there is only $SU(2)$ spin rotation symmetry and the singlon and doublon carry $S=1/2$ and $S=1$ respectively.  This t-J model can be derived from the $SU(4)$ symmetric $t-J$ model in this paper by adding anisotropic terms. Therefore our analysis in this paper may also provide insights to the solid-state realization of type II $t-J$ model using the two $e_g$ orbitals.

In the familiar $t-J$ model, at least for large doping, the most natural ground state is a conventional Fermi liquid.  This state can be constructed within the simple slave-boson mean field theory\cite{lee2006doping} which respects the constraint $P_1$. The simple way to understand this Fermi liquid state is that the holons condense and the singlons form Fermi surfaces.  This picture can be easily generalized to the case $\nu_T=1-x$ for the spin-valley Hubbard model. However, for $\nu_T=2-x$, neither the singlons nor the doublons can form Fermi surfaces whose area match a conventional Fermi liquid. In this case, description of a conventional Fermi liquid is very hard if we insist to respect the constraint $P_s+P_d$. As we show later, a generalized slave-boson mean field theory naturally predicts pseudogap metals with Fermi surface areas proportional to $x$ instead of $\nu_T$. We can still describe a conventional Fermi liquid phase, but it requires a more exotic parton construction if we want to respect the constraint $P_2$.

\section{Symmetry constraint: Luttinger theorem}
Before we move to discussions of specific fillings, we give a general symmetry analysis in this section. We will consider LSM type of constraints based on simple Oshikawa-flux threading argument\cite{oshikawa2000topological}. The argument and the constraint works for both integer and incommensurate filling. Besides, the symmetry analysis in this section is independent of models and also applies to the case with topological bands.

The symmetries that we consider here are translation, spin rotation, charge conservation and time reversal. Depending on whether we include the inter-valley spin-spin coupling, we consider $U(2)_+\times U(2)_-$ spin rotation symmetry and $U(1)_c \times U(1)_v \times SO(3)_s$ separately. In all cases we assume there is a time reversal symmetry which exchanges the two valleys.

\subsection{Symmetry $U(4)$ or $U(2)_+\times U(2)_-$}
The constraint is the same for $U(4)$ and $U(2)_+\times U(2)_-$. For simplicity we will only use $U(2)_+\times U(2)_-$. Note that time reversal exchanges two valleys so the density for each flavor is guaranteed to be $\nu=\frac{\nu_T}{4}$. Meanwhile we have $U(1)^4 \subset U(2)_+\times U(2)_-$, which means we have $U(1)$ symmetry for each flavor. Then, we can do flux insertion for one valley-spin species out of four.  Using Oshikawa's argument\cite{oshikawa2000topological}, one can reach the conclusion that any symmetric and featureless phase needs to have Fermi surface area $A_{FS}=\nu \mod n=\frac{\nu_T}{4} \mod n$, where $n$ is an integer. 

For $\nu_T=1,2$, the above constraint  means there must be a Fermi surface with area $1/4$ or $1/2$ for each flavor. Therefore symmetric and featureless Mott insulator is not possible for symmetry $U(2)_+\times U(2)_-$. 

\subsection{Symmetry $(U(1)_c\times U(1)_v \times SU(2)_S))/Z_2$: Two distinct symmetric and featureless states}

When there is a non-zero $J_H$, the global symmetry is $(U(1)_c\times U(1)_v \times SU(2)_S))/Z_2$. There is no separate spin rotation symmetry for each valley. Only the total spin rotation is conserved in this case. There are only three independent $U(1)$ global symmetries(corresponding to $N$, $S_z$ and $\tau_z$) that belong to $(U(1)_c\times U(1)_v \times SU(2)_S)/Z_2$ (Recall that there are four $U(1)$ charges in the previous subsection.). Gauging these three $U(1)$ symmetries yields three independent  flux insertions. we cannot do flux insertion for each valley-spin species. The best we can do is to include at least two valley-spin species in the flux insertion process in order to respect the global symmetry. Since we still have time reversal and total spin rotation symmetry, the filling per valley per spin is still fixed to be $\nu=\frac{\nu_T}{4}$, $\nu_T$ being the total filling. If the ground state is a symmetric Fermi liquid, the Fermi surface area for every flavor must be equal to each other.  Let us perform an adiabatic $U(1)$ flux insertion for both spins of the $+$ valley. The constraint one can get using Oshikawa's argument is $A_{FS,+\uparrow}+A_{FS,+\downarrow} = 2A_{FS}=\nu_{+\uparrow}+\nu_{+\downarrow} (\mod~n)=\frac{\nu_T}{2} (\mod~n) $, where $n$ is an integer. This yields $A_{FS} = \frac{\nu_T}{4} (mod~m)$ or $A_{FS} =( \frac{\nu_T}{4}-\frac{1}{2} )(mod~m)$, where $m$ is an integer.

Interestingly, we find that there are two distinct symmetric and featureless Fermi liquids. One of them is connected to the free fermion model while the other one has smaller Fermi surfaces and may be viewed as a "pseudogap metal". The essential point of the argument is that there are only three $U(1)$ charges while there are four flavors. Meanwhile, the symmetry is sufficient to forbid bilinear term with inter-flavor coupling and guarantee four equal Fermi surfaces. The above two conditions can not be satisfied simultaneously in the traditional spin $1/2$ system.  This is a special feature of spin-valley model realized in  moir\'e superlattices.

  At $\nu_T=2$, we are allowed to have $A_{FS}=1/2$ or $A_{FS}=0$, which implies the existence of a symmetric and featureless Mott insulator.  We need to emphasize that $U(1)_v$ and time reversal symmetry is important to guarantee the existence of the two distinct class. Without time reversal, we can have a "band insulator" at $\nu_T=2$ by polarizing valley. This band insulator can smoothly connect to the conventional Fermi liquid by reducing the $\tau_z$ polarization.  Once we have time reversal symmetry, the density of each flavor is fixed to be $1/2$ at $\nu_T=2$. In this case the featureless Mott insulator can not smoothly cross over to the conventional Fermi liquid and can not be described by mean field theory with any order parameter.  

We will construct both the featureless insulator at $\nu_T=2$ and the featureless pseudogap metal in Section V and Section VII for the spin-valley Hubbard model. The essential physics behind them is \textbf{singlet formation}. At $\nu_T=2$,  there are N electrons in the valley $+$ and $N$ electrons in the valley $-$, where $N$ is the number of moir\'e unit cells. These electrons can be gapped out by forming inter-valley singlets.  If we further dope electrons or holes with density $x$, these additional doped carriers just form small Fermi surfaces with area $A_{FS}=\frac{x}{4}$ on top of the featureless Mott insulator.

Although the picture of the featureless insulator and the featureless pseudogap metal is very simple, they can not be described by the conventional mean field theory with symmetry breaking order parameters. In the cuprate context, symmetric pseudogap metals with small Fermi surfaces have been proposed before\cite{senthil2003fractionalized,sachdev2016novel}. In the so called FL* phase, additional holes form small Fermi surfaces on top of a $Z_2$ spin liquid. The physics behind is still singlet formation: $N$ number of electrons form resonating valence bond (RVB) singlets and the additional holes move on top of the RVB singlets. The difference in our case is that we can have on-site inter-valley singlet and do not need to invoke fractionalization. In this sense, the featureless pseudogap metal we propose here is the simplest version of symmetric pseudogap metal.

This simple symmetric pseudogap metal is beyond any conventional mean field  theories with symmetry breaking because the singlet formation does not break any symmetry. So how do we describe the singlet formation? It turns out that the singlet formation can be captured in a simple six-flavor slave boson parton mean field theory.  We will introduce the parton construction for the $\nu_T=2$ Mott insulator first and then generalize it to the doped case to describe the featureless pseudogap metal. This slave boson parton also allow us to describe another orthogonal pseudogap metal in the $SU(4)$ symmetric limit.

\onecolumngrid

\begin{table}[H]
\centering
\begin{tabular}{|c|c|c|c|c|c|c|c|c|c|c|c|c|c|c|}
\hline
$\tau_x$&$\tau_y$&$\tau_z$ & $\sigma_x$ &$\sigma_y$ &$\sigma_z$ &$\tau_x \sigma_x$ & $\tau_x \sigma_y$ & $\tau_x \sigma_z$&$\tau_y \sigma_x$ & $\tau_y \sigma_y$ & $\tau_y \sigma_z$ &$\tau_z \sigma_x$ & $\tau_z \sigma_y$ & $\tau_z \sigma_z$\\
\hline
$S_{32}$&$S_{31}$&$S_{12}$&$S_{64}$& $S_{45}$&$S_{65}$ &$S_{15}$ &$S_{16}$ &$S_{41}$&$S_{52}$ &$S_{62}$ &$S_{24}$ &$S_{53}$ &$S_{63}$ &$S_{34}$\\
\hline
\end{tabular}
\caption{The correspondence between the generator of the $SO(6)$ and the generators of the $SU(4)$. For example, the $SU(4)$ transformation $U=e^{i \tau_z \frac{\theta}{2}}$ corresponds to a rotation between $\psi_1$ and $\psi_2$ with angle $\theta$.}
\label{table:so(6)_generators}
\end{table}

\twocolumngrid

\section{Projective Symmetry Group analysis at $\nu_T=2$}

\subsection{Hilbert Space}
At $\nu_T=2$, the Hilbert space is six dimensional at each site. There are six bases:  $\ket{\alpha}=\sum_{ab} A^\alpha_{ab} c^\dagger_a c^\dagger_b \ket{0}$ with $\alpha=1,2,3,4,5,6$.  $A^\alpha$ is a $4\times 4$ anti-symmetric matrix.

We define the $SU(4)$ flavor as $1: +\uparrow$, $2: +\downarrow$, $3: -\uparrow$, $4 -\downarrow$.  Each basis $\alpha$ is created by an operator $\Psi_\alpha^\dagger$.  We can define the following six bases:
\begin{align}
	\psi_1&=\frac{1}{2\sqrt{2}}c^T  \tau_z \sigma_y c=\frac{i}{\sqrt{2}}(-\Phi_{12}+\Phi_{34})\notag\\
	\psi_2&=\frac{1}{2\sqrt{2}}c^T  i \sigma_y c=\frac{1}{\sqrt{2}}(\Phi_{12}+\Phi_{34})\notag\\
	\psi_3&=\frac{1}{2\sqrt{2}}c^T  \tau_x \sigma_y c=\frac{i}{\sqrt{2}}(-\Phi_{14}+\Phi_{23})\notag\\
	\psi_4&=\frac{1}{2\sqrt{2}}c^T  i\tau_y \sigma_x c=\frac{1}{\sqrt{2}}(\Phi_{14}+\Phi_{23})\notag\\
	\psi_5&=\frac{1}{2\sqrt{2}}c^T  i\tau_y  \sigma_z c=\frac{1}{\sqrt{2}}(\Phi_{13}-\Phi_{24})\notag\\
	\psi_6&=\frac{1}{2\sqrt{2}}c^T  \tau_y c=-\frac{i}{\sqrt{2}}(\Phi_{13}+\Phi_{24})\notag\\
\label{eq:SO(6)_rep_obvious_basis}
\end{align}
where $\Phi_{ab}=\frac{1}{2}(c_ac_b-c_bc_a)$. 

These six states are organized to have clear physical meaning: the first three are valley triplet and spin singlet while  the later three are valley singlet and spin triplet.

Let us define $\Psi^T=(\psi_1,\psi_2,\psi_3,\psi_4,\psi_5,\psi_6)$. It can be proved that the transformation is $\Psi'=O \Psi$ under the microscopic $SU(4)$ transformation. $O$ is a $SO(6)$ matrix.  There are $15$ generators for the $SO(6)$. We list them in Table.~\ref{table:so(6)_generators}.

The physical spin operator $S^{\mu \nu}$ defined in Eq.~\ref{eq:spin_operator_definition} can be written as a $6 \times 6$ matrix in the above basis. It is convenient to express it as $S^{\mu \nu}=\Psi^\dagger A^{\mu \nu} \Psi$. It turns out that the $6 \times 6$ matrix $A^{\mu \nu}$ only has two non-zero matrix elements.  More specifically, each spin operator $S^{\mu \nu}$ corresponds to a $(\alpha,\beta)$ pair as listed in Table.~\ref{table:so(6)_generators}.  Then $S^{\mu\nu}=S^{\alpha \beta}=2i (\psi^\dagger_{\alpha}\psi_{\beta}-\psi^\dagger_{\beta} \psi_\alpha)$. For example, $\tau^x=2i(\psi^\dagger_3\psi_2-\psi^\dagger_2\psi_3)$.

When $H'_S=H''_S=0$, the spin model has $SO(6)/Z_2$ symmetry where $Z_2$ consists of the $6\times 6$ matrix $- I$. When $H'_S\neq 0$, the spin rotation symmetry is $(SO(2)\times SO(4))/Z_2$.  The Hilbert space at each site is decomposed to $6 = 2 \bigoplus 4 $. $(\psi_1,\psi_2)$ transforms as $SO(2)$ corresponding to $U(1)_{valley}$. $(\psi_3,\psi_4,\psi_5,\psi_6)$ transforms as $SO(4)$ under $SU(2)_+\times SU(2)_-$.  If we further add the on-site inter-valley spin coupling $H''_S$, the spin rotation symmetry is further reduced to $SO(2)\times SO(3)$.  The Hilbert space at each site is decomposed to three irreducible representations: $6=1\bigoplus 2 \bigoplus 3$. $\psi_3$ forms a trivial representation. $(\psi_1,\psi_2)$ transforms under $SO(2)$ and $(\psi_4,\psi_5,\psi_6)$ transforms under $SO(3)$ in the same way as a spin-one degree of freedom.

For spin rotation symmetry $SO(6)/Z_2$ and $(SO(2)\times SO(4))/Z_2$, there is no symmetric gapped state. For the symmetry $SO(2)\times SO(3)$, a featureless insulator is possible. For example, the state $\prod_i \psi_3^\dagger (i) \ket{0}$ is a featureless Mott insulator. However, we need  a large $J_H>0$ to favor this featureless insulating phase. In the trilayer graphene/h-BN system, we expect $|J_H|$ to be smaller or at most comparable with $J$.  When $J>>|J_H|$, other phases including magnetic order, valence bond solid (VBS) and spin liquids may be favored. In this paper we try to classify all of possible symmetry spin liquid phases based on PSG analysis.

\subsection{Parton Theories  at $\nu_T=2$}
For $\nu_T=2$, there are three different parton theories which we introduce in this subsection.

\subsubsection{Abrikosov Fermion}
The first parton theory is the Abrikosov fermion parton which has been widely used in the treatment of spin $1/2$ systems. We introduce fermionic operator $f_{i;a}$ with $a=1,2,3,4$. The constraint is $\sum_a f^\dagger_{i;a}f_{i;a}=2$. There is a $U(1)$  gauge structure. As we will discuss later, the only symmetric spin liquid from the Abrikosov fermion parton is $U(1)$ spin liquid with $4$ Fermi surface, each at filling $1/2$.

\subsubsection{Six-flavor Schwinger Boson}
Because the Hilbert space at each site is six dimensional and form the fundamental representation of $SO(6)$, we can use a six flavor Schwinger boson parton construction. Basically we identify $\Psi_\alpha$ in Eq.~\ref{eq:SO(6)_rep_obvious_basis} as a bosonic operator $b_{\alpha}$, where $\alpha=1,2,...,6$. Correspondingly the spin operator $S^{\alpha \beta}=2i (b^\dagger_{\alpha} b_{\beta}-b^\dagger_{\beta}b_{\alpha})$. For simplicity, we define a six dimensional spinor $B(\mathbf r)=(b_1(\mathbf r),b_2(\mathbf r),b_3(\mathbf r),b_4 (\mathbf r),b_5 (\mathbf r),b_6(\mathbf r))$. The constraint is $B^\dagger(\mathbf r)B(\mathbf r)=1$ for each site $\mathbf r$. The gauge structure is $U(1)$.

We define hopping term $\hat T_{ij}=B^\dagger_i B_j$ and $\hat \Delta_{ij}=\frac{1}{\sqrt{2}}(B_i^T B_j+B_j^T B_i)$. $\hat T^\dagger_{ij}=T_{ji}$. $\hat \Delta^\dagger_{ij}=\frac{1}{\sqrt{2}}(B^{\dagger}_j B^{\dagger T}_i+B_i^\dagger B^{\dagger T}_j)$. Apparently $\hat \Delta_{ij}=\hat \Delta_{ji}$.

The Hamiltonian in the $SO(6)$ invariant limit can be written as:
\begin{equation}
	H=4J\sum_{\langle ij \rangle}\hat T_{ij} \hat T_{ji} -4J \sum_{\langle ij \rangle} \hat \Delta^\dagger_{ij} \hat \Delta_{ij}
\end{equation}

In the Schwinger boson theory, typical mean field ansatz is:
\begin{equation}
	H_M=\sum_{\langle ij \rangle} \left(T^\dagger_{ji} \hat T_{ij}+ \Delta^\dagger_{ji} \hat{\Delta}_{ij} \right)
\end{equation}
where $T_{ij}$ and $\Delta_{ij}$ are mean field parameters. 

For Schwinger boson, meaningful ansatz has $\Delta_{ij} \neq 0$, which describes a $Z_2$ spin liquid.

\subsubsection{ Six-flavor Schwinger Fermion}
Similar to the  six flavor Schwinger boson approach, we can also identify each basis $\ket{\alpha}$ in Eq.~\ref{eq:SO(6)_rep_obvious_basis} to be created by a fermionic operator $\Psi_{\alpha}$. Defining $\Psi=(\Psi_1,\Psi_2,\Psi_3,\Psi_4,\Psi_5,\Psi_6)$. The constraint is $\Psi^\dagger(\mathbf r)\Psi(\mathbf r)=1$ for each site $\mathbf r$. The gauge structure is $O(2)=Z_2 \rtimes U(1)$. $Z_2$ corresponds to a charge conjugation $C: \Psi(\mathbf r)\rightarrow \Psi^\dagger(\mathbf r)$.

We define hopping term $\hat T_{ij}=\Psi^\dagger_i \Psi_j$ and $\hat \Delta_{ij}=\frac{1}{2}(\Psi_i \Psi_j-\Psi_j\Psi_i)$. $\hat T^\dagger_{ij}=\hat T_{ji}$. $\hat \Delta^\dagger_{ij}=\frac{1}{2}(\Psi^\dagger_j \Psi^\dagger_i-\Psi_i^\dagger \Psi^\dagger_j)$. Apparently $\hat \Delta_{ij}=-\hat \Delta_{ji}$.

The  $SO(6)$ invarinat Hamiltonian  can be written as
\begin{align}
	H&=-4J\sum_{\langle ij \rangle}\hat T_{ij} \hat T_{ji} -4 J \sum_{\langle ij \rangle} \hat \Delta^\dagger_{ij} \hat \Delta_{ij}\notag\\
\end{align}

Typical mean field ansatz in the Schwinger fermion approach is:

\begin{equation}
	H_M=\sum_{\langle ij \rangle} \left(T^\dagger_{ji} \hat T_{ij}+ \Delta^\dagger_{ji} \hat{\Delta}_{ij} \right)
\end{equation}
where $T_{ij}$ and $\Delta_{ij}$ are mean field parameters.

For Schwinger fermion parton theory, we can have both $U(1)$ spin liquid with $\Delta_{ij}=0$ and $Z_2$ spin liquid with $\Delta_{ij}\neq 0$. If $\Delta_{ij}$ is chiral, we can also have chiral spin liquid.

\subsection{PSG Classification of $U(1)$ Spin Liquids at $\nu_T=2$}

 We first discuss $U(1)$ spin liquid. There are two different kinds of $U(1)$ spin liquid phases. One is constructed from the Abrikosov fermion and the other one is constructed from the Schwinger fermion. For both Abrikosov fermion and Schwinger fermion, symmetry $U(1)$ spin liquids have the same classification as shown in Appendix.~\ref{appendix: U(1) PSG}.  Both zero-flux phase and $\pi$-flux phase are symmetric. However, nearest neighbor and next nearest neighbor hopping in the $\pi$-flux phase are forbidden by the PSG, which is not physical.  Therefore we only consider the zero flux phase for both the Abrikosov fermion parton and  the Schwinger fermion parton.

From Abrikosov fermion, we have a $U(1)$ spin liquid with four Fermi surfaces, each of which is at filling $1/2$. This spin liquid phase can go through a continuous transition to Fermi liquid. Basically one can write the electron operator as $c_{i;a}={e^{i\theta_i} f_{i;a}}$ and let the bosonic rotor go through a continuous superfluid-Mott transition\cite{senthil2008theory}.

From the Schwinger fermion, we have a $U(1)$ spin liquid with six Fermi surfaces, each of which is at filling $1/6$. This $U(1)$ spinon fermi surface phase is  not connected to the Fermi liquid through direct transition.

\subsection{PSG Classification of $Z_2$ Spin Liquids at $\nu_T=2$}

We then classify symmetric $Z_2$ spin liquids at $\nu_T=2$. $Z_2$ spin liquid can be constructed from both the Schwinger boson parton and the Schwinger fermion parton.  The PSG is the same for both Schwinger boson and  Schwinger fermion parton constructions. It is independent of the spin rotation symmetry and therefore is true even the $SO(6)/Z_2$ spin rotation is broken down to $SO(2)\times SO(3)$.

 PSG classification is the same as the spin $1/2$ Schwinger boson approach, and we can just quote the results of Wang et.al. in Ref.~\onlinecite{wang2006spin}.  The PSG is labeled by $(p_1,p_2,p_3)$ where $p_1,p_2,p_3=0,1$.  These three integers label the fractionalization of the translation, $\sigma$ and $C_6$: $T_1T_2=T_2 T_1 (-1)^{p_1}$, $\sigma^2=(-1)^{p_2}$ and $C_6^6=(-1)^{p_3}$. 

For each symmetry operation $X$, the corresponding projective symmetry operation is $G_X X$ where $G_X=e^{i \varphi_X(\mathbf r)}$ is a $U(1)$ gauge transformation.  
\begin{align}
&\varphi_{T_1}(x,y)=0\notag\\
&\varphi_{T_2}(x,y)=p_1 \pi x\notag\\
&\varphi_{\sigma}(x,y)=p_2\frac{\pi}{2}+p_1 \pi xy\notag\\
&\varphi_{C_6}(x,y)=p_3 \frac{\pi}{2}+p_1 \pi(xy+\frac{y(y-1)}{2})
\end{align}
where the coordinate of a site is $\mathbf{R}=x \mathbf{a_1}+y \mathbf{a_2}$.  The transformation of crystal symmetries can be found in Appendix.~\ref{append:lattice_symmetries}.

For both Schwinger boson and Schwinger fermion, there are 8 different $Z_2$ Spin liquids labeled by $(p_1,p_2,p_3)$. However, more constraints need to be added if we require the nearest neighbor pairing to be non-zero. This is a reasonable requirement for a model with dominant nearest-neighbor anti-ferromagnetic coupling. It turns out there are only two $Z_2$ Spin liquids for both fermion and boson parton in the SO(6) symmetric limit. In the following we will introduce the two PSG ansatz for the Schwinger boson and Schwinger fermion respectively. Then we will connect the Schwinger fermion and Schwinger boson approaches and show that they describe the same $Z_2$ spin liquids in terms of $f$ particle and $e$ particle.

\subsubsection{$Z_2$ Spin Liquid in Schwinger Fermion Parton}
The only $SO(6)$ symmetric pairing is $\Delta(\mathbf R) \sum_a\Psi_a (\mathbf r) \Psi_a(\mathbf r+\mathbf{R})$ where $a=1,...,6$ is the flavor index. For fermion we have $\Delta(\mathbf R)=-\Delta(-\mathbf R)$.  We require $\Delta(\mathbf R)\neq 0$ for nearest neighbor $\mathbf R$.

First, the reflection $\sigma$ maps $\mathbf R=(1,1)$ to itself.  Under the PSG, $\Delta(1,1)\rightarrow \Delta(1,1)e^{i(\varphi_\sigma(0,0)+\varphi_\sigma(1,1))}$.  To have $\Delta(1,1)\neq 0$, we need  $\varphi_\sigma(0,0)+\varphi_\sigma(1,1))=0\  mod \ 2\pi $.  This is equivalent to $p_1=p_2$.

Second, $\mathbf{R_2}=(-1,0)$ is related to $\mathbf{R_1}=(1,0)$ by three $C_6$ rotation. Meanwhile we have $\Delta(-1,0)=-\Delta(1,0)$. Therefore we have the following condition:
\begin{equation}
	3 \varphi_{C_6}(0,0)+\varphi_{C_6}(1,1)+\varphi_{C_6}(0,1)+\varphi_{C_6}(-1,0)= \pi \ mod\ 2\pi
\end{equation}
which requires $p_3=1-p_1 \ mod\ 2$.

Because of the above two constraints, there are only two  reasonable symmetric $Z_2$ spin liquids with $SO(6)$ rotation symmetry. They are $(0,0,1)$ and $(1,1,0)$ phase. The first is a zero flux phase while the second is a $\pi$ flux phase.

Next we show that $p_1 =1$ forbids nearest neighbor hopping.  Consider the hopping $t(\mathbf R)\sum_a \Psi_a^\dagger(\mathbf{r+R})\Psi_a(\mathbf r)$. $\mathbf{R_1}=(0,1)$ and $\mathbf{R_2}=(1,0)$ are related by $\sigma$ or $C_6^2$. To have non-zero nearest neighbor hopping, we need to have:
\begin{equation}
	-\varphi_\sigma(0,1)+\varphi_\sigma(0,0)=-\varphi_{C_6}(1,1)-\varphi_{C_6}(0,1)+2\varphi_{C_6}(0,0) \ mod \ 2\pi
\end{equation}
which fixes $p_1=0$.

 As a result, the $\pi$ flux phase needs to have zero nearest neighbor hopping.

\subsubsection{$Z_2$ Spin Liquid in Schwinger Boson Parton}
For six flavor boson,  the pairing is also $\Delta(\mathbf R) \sum_a\Psi_a (\mathbf r) \Psi_a(\mathbf r+\mathbf{R})$ for each bond $\mathbf R$. We have $\Delta(\mathbf R)=\Delta(-\mathbf R)$.  Similar to the fermion case, the reflection $\sigma$ fixes $p_1=p_2$.  The $C_6^3$ relates $\mathbf{R_1}=(1,0)$ to $\mathbf{R_2}=(-1,0)$ and fixes $p_3=p_1$.

There are also two symmetric spin liquids satisfying the constraint: the zero flux phase $(0,0,0)$ and  the $\pi$ flux phase $(1,1,1)$. Again the $\pi$ flux phase can not have non-zero nearest neighbor hopping.

\subsubsection{Equivalence between Schwinger Boson and Schwinger Fermion Description}
We have found two  symmetric $Z_2$ spin liquids from both Schwinger boson and Schwinger fermion construction. In this section we show that the Schwinger fermion descriptions are equivalent to the Schwinger boson approach.

In the Schwinger boson approach, we have the PSG for the bosonic $e$ particle. In the Schwinger fermion parton theory, we have the PSG for the fermionic $f=e \upsilon $. Here $\upsilon$ denotes the vison or $m$ particle. The PSG of $f$ particle can be derived from the composition of the PSG of the $e$ and $m$ particle (with possible twisting factor)\cite{essin2013classifying,lu2017unification,qi2015detecting}.  There is one $e$ particle per unit cell and the vison $\upsilon$ always see the $e$ particle as an effective $\pi$ flux. Thus the vison always has the PSG $T_1 T_2=-T_2 T_1$ and $C_6^6=-1$. It has been shown that $\sigma^2=-1$ is anomalous for the vision when there is a $U(1)$ spin rotation symmetry. Vison must have $\sigma^2=1$ \cite{hermele2016flux} in our problem.  For the fractionalization of $X=T_1 T_2 (T_2 T_1)^{-1}, \sigma^2, C_6^6$,  the PSG of vison $(-1)^{p^\upsilon_X}$ can be uniquely determined as $P^\upsilon=(-1,0,-1)$. We can then get PSG of $f$ from the PSG of $e$ particle by: $(-1)^{P^f_X}=(-1)^{P^e_X}(-1)^{P^\upsilon_X} (-1)^{\epsilon(X)}$, where $\epsilon(X)=1,-1$ is a twist factor. It has been shown that $\epsilon(T_1 T_2 (T_2 T_1)^{-1})=1$ and $\epsilon(\sigma^2)=\epsilon(C_6^6)=-1$\cite{lu2017unification,qi2015detecting}.  We can then map PSG of $e$ particle $P^b=(p^b_1,p^b_2,p^b_3)$ to $P^f=(p^f_1,p^f_2,p^f_3)$ by equation:
\begin{equation}
	p^f_1=p^b_1+1,\  p^f_2=p^b_2+1,\  p^f_3=p^b_3
\end{equation}

From the above relation we can see that both Schwinger boson theory and Schwinger fermion theory are describing  two symmetric $Z_2$ spin liquids: I. $P^b=(0,0,0)$ and $P^f=(1,1,0)$. II. $P^b=(1,1,1)$ and $P^f=(0,0,1)$.

In summary we find two symmetric $Z_2$ spin liquid,shown in Table.~\ref{table:Z2_so_6}.  Each of them can be described using either Schwinger boson or Schwinger fermion mean field ansatz. Details about the ansatz and the dispersion can be found in the Appendix.~\ref{append:Z_2_spin_liquid}.

\begin{table}[H]
\centering
\begin{tabular}{c|c|c|c}

\hline
Phase&PSG& Band Bottom of $e$  \\
\hline
Type I $Z_2$ & $P^b=(0,0,0)$, $P^f=(1,1,0)$  & $(0,0)$ \\
\hline
Type II $Z_2$ & $P^b=(1,1,1)$, $P^f=(0,0,1)$ & $\pm(\frac{\pi}{6},\frac{\pi}{2\sqrt{3}})$,$\pm(\frac{5\pi}{6},\frac{\pi}{2\sqrt{3}})$ \\
\hline
\end{tabular}
\caption{Two symmetric $Z_2$ spin liquids. $P^b$ and $P^f$ label symmetric fractionalization for $e$ and $f$ particles. We also list the band bottom of $e$ particle in the Schwinger boson mean field ansatz with only nearest neighbor coupling.}
\label{table:Z2_so_6}
\end{table}

\section{Pseudogap Metals at $\nu_T=2-x$}
After discussing the Mott insulator, we turn to possible metallic phases upon doping the Mott insulator. Especially, we show that pseudogap metals with small Fermi surfaces can naturally emerge upon doping the Mott insulator at $\nu_T=2$. 

The Mott insulating phase at $\nu_T=2$ is sensitive to the on-site inter-valley spin-spin coupling $J_H$. A non-zero $J_H$ split the $SO(6)$ symmetry down to $1\oplus 2 \oplus 3$. When $J_H>0$ and its magnitude is  much larger than $J$, it is obvious that the ground state is a featureless Mott insulator with one valley triplet, spin singlet at each site. When $J_H<0$ with a large magnitude, the low energy is dominated by one $SO(3)$ vector at each site. Therefore we have an effective spin-one model on triangular lattice and the ground state is the $120^\circ$ Neel order. For the intermediate region with $J_H=0$, valence bond solid (VBS) or resonant valence bond (RVB) may also be possible.

In the remaining part we discuss possible metallic phases upon doping from the featureless Mott insulator and the VBS/RVB phase.  Physics of doping the Neel order is very hard and we leave it to future work.

\subsection{$J_H>0$: Symmetric and featureless pseudogap Metal}
In the simplest case, let us assume $J_H>0$ and is much larger than $J$. In this case the Mott insulator has one inter-valley singlet at each site. When we dope the system to the filling $\nu_T=2-x$, there are $x$ singlons. One natural state is that these singlons move coherently and form four Fermi surfaces, each of which has area $A_{FS}=-\frac{x}{4}$. The remaining particles are still gapped out by singlet formation. Obviously this is a pseudogap metal with only a small Fermi surface. The Hall number is opposite to the free fermion case and is proportional to $x$. This is quite similar to the phenomenology of cuprates when superconductivity is suppressed by strong magnetic field. This pseudogap metal is a symmetric Fermi liquid. Although the picture is very simple, this state is definitely beyond the conventional scenario with density wave order. The existence of this simple example is a proof that electrons can be gapped out from the Fermi sea without invoking any symmetry breaking order.

Although we can not describe this featureless pseudogap phase with the conventional mean field theory, we find that the essential physics can be easily captured by a slave boson mean field theory. When we remove one electron from $\nu_T=2$, we remove one doublon and create one singlon, therefore we use the following parton representation:
\begin{equation}
	c_{\alpha}(\mathbf x)=\sum_{\beta \neq \alpha} \Phi_{\alpha \beta}(\mathbf x) f^\dagger_\beta(\mathbf x)
	\label{eq:six_flavor_parton_electron}
\end{equation}
where $\alpha,\beta=1,2,3,4$ is the flavor index. $\Phi_{\alpha \beta}(\mathbf{x})=-\Phi_{\beta \alpha}(\mathbf{x})$ is an anti-symmetric slave boson field, which has been used before in Eq.~\ref{eq:SO(6)_rep_obvious_basis}. $f^\dagger_\beta(\mathbf x)$ creates a singlon with flavor $\beta$.  The above parton construction has a $U(1)$ redundancy:
\begin{equation}
	\Phi_{\alpha \beta}(\mathbf x)\rightarrow \Phi_{\alpha \beta}(\mathbf x)e^{i\theta(\mathbf x)}\ \ \  f_{\beta}(\mathbf x)\rightarrow f_{\beta}(\mathbf x)e^{i \theta(\mathbf x)}
\end{equation}
with the constraint $n_f(\mathbf x)+n_b(\mathbf x)=1$.

When $J_H=0$, these six $\Phi_{\alpha \beta}$ fields can form a $SO(6)$ vector in the basis defined in Eq.~\ref{eq:SO(6)_rep_obvious_basis}. In the following we use the $SO(6)$ basis $b_a(\mathbf x)$ with $a=1,2,...,6$:

\begin{align}
	b_1&=\frac{i}{\sqrt{2}}(-\Phi_{12}+\Phi_{34})\notag\\
	b_2&=\frac{1}{\sqrt{2}}(\Phi_{12}+\Phi_{34})\notag\\
	b_3&=\frac{i}{\sqrt{2}}(-\Phi_{14}+\Phi_{23})\notag\\
	b_4&=\frac{1}{\sqrt{2}}(\Phi_{14}+\Phi_{23})\notag\\
	b_5&=\frac{1}{\sqrt{2}}(\Phi_{13}-\Phi_{24})\notag\\
	b_6&=-\frac{i}{\sqrt{2}}(\Phi_{13}+\Phi_{24})\notag\\
\label{eq:SO(6)_slave_boson_basis}
\end{align}

  We can substitute Eq.~\ref{eq:SO(6)_slave_boson_basis} to  the $t-J$ model in Eq.~\ref{eq:unconventional_t_J} and decouple it in the form of mean field theory:
\begin{align}
H_M&=H_b+H_f\notag\\
H_b&=-t_b \sum_{\langle ij \rangle}\sum_a b^\dagger_{a;i} b_{a;j}+h.c.-\mu \sum_i \sum_a b^\dagger_{a;i}b_{a;i}\notag\\
H_f&=-t_f \sum_{\langle ij \rangle}\sum_\alpha f^\dagger_{\alpha;i}f_{\alpha;j}+h.c.-\mu_f \sum_i \sum_\alpha f^\dagger_{\alpha;i}f_{\alpha;i}
\label{eq:six_flavor_meanfield}
\end{align}

When we add a $J_H>0$, the degeneracy of the six flavor bosons is lifted and the one corresponding to the inter-valley singlet is favored. This is $b_3=\frac{i}{\sqrt{2}}(-\Phi_{14}+\Phi_{23})$  defined in Eq.~\ref{eq:SO(6)_slave_boson_basis}. Therefore we consider the ansatz with $\langle b_3(\mathbf x) \rangle \neq 0$. After the condensation of $b$, the internal $U(1)$ gauge field is higgsed and $f$ can be identified as a local hole operator $c^\dagger$. Because the density of fermion $f$ is $n_f=x$. We have four Fermi surfaces with Fermi surface area $A_{FS}=-\frac{x}{4}$.

To further prove the phase is a Fermi liquid, we can try to calculate the single Green function. With simple algebras, we can easily get:
\begin{align}
	&G^c_{\alpha_1,\alpha_2}(x,t;y,t')=\langle c^\dagger_{\alpha_1}(x,t) c_{\alpha_2}(y,t') \rangle \notag\\
	&= \sum_{\beta_1\neq \alpha_1}\sum_{\beta_2 \neq \alpha_2}\langle \Phi^\dagger_{\alpha_1 \beta_1}(x,t) \Phi_{\alpha_2 \beta_2}(y,t') \rangle \langle f_{\beta_1}(x,t)f^\dagger_{\beta_2}(y,t')\rangle \notag\\
	&=\sum_{\beta \neq \alpha_1, \alpha_2} \langle \Phi^\dagger_{\alpha_1 \beta}(x,t) \Phi_{\alpha_2 \beta}(y,t') \rangle  G^f_\beta(x,t;y,t')
\end{align}

Using the fact that $\langle \Phi_{14} \rangle=-\langle \Phi_{23} \rangle=\frac{i}{\sqrt{2}}\langle b_4 \rangle$ while other components of $\Phi_{\alpha \beta}$ is zero, we can easily get
\begin{align}
G^c_{11}(x,t;y,t')&=\frac{1}{2}|\langle b \rangle|^2 G^f_{44}(x,t;y,t')\notag\\
G^c_{22}(x,t;y,t')&=\frac{1}{2}|\langle b \rangle|^2 G^f_{33}(x,t;y,t')\notag\\
G^c_{33}(x,t;y,t')&=\frac{1}{2}|\langle b \rangle|^2 G^f_{22}(x,t;y,t')\notag\\
G^c_{44}(x,t;y,t')&=\frac{1}{2}|\langle b \rangle|^2 G^f_{11}(x,t;y,t')\notag\\
\end{align}
These equations also mean that $f$ operator should be identified as a physical hole operator.

In summary, in the limit with a large anti inter-valley Hund's coupling, a natural state in the under-doped regime is a symmetric and featureless Fermi liquid with small Fermi surfaces. Such a state has Hall number $\eta_H=-x$ for $\nu_T=2-x$, in contrast to the free fermion case with $\eta_H=2-x$. This simple state offers a simple example to partially gap out free fermion Fermi surfaces by symmetric singlet formation, instead of the more familiar density wave order.

\subsection{$J_H=0$: orthogonal metal with small Fermi surfaces}

Next we turn to the case with $J_H=0$. In this $U(4)$ symmetric point, Luttinger theorem requires $A_{FS}=\frac{\nu_T}{4}$ for a symmetric and featureless phase. However, we will argue that an exotic symmetric pseudogap metal may be possible when doping away from the $\nu_T=2$ Mott insulator in the $U>>t$ limit. We will generalize the conventional RVB theory\cite{lee2006doping} to the type II $t-J$ model close to $\nu_T=2$.  As in the familiar RVB theory, we assume the undoped state is a $Z_2$ spin liquid.

We still use the parton construction in Eq.~\ref{eq:six_flavor_parton_electron}. We have $n_f=x$ and $n_b=1-x$. In the undoped Mott insulator, there are two $Z_2$ spin liquids as we proposed before. For simplicity we just use the zero flux ansatz from the Schwinger boson method. Basically, the schwinger boson $b$ is in a paired superfluid phase. Then we dope the system, $f$ can form four Fermi surfaces with area $A_{FS}=-\frac{x}{4}$ as in the featureless pseudogap metal in the previous subsection.  In this mean field ansatz, $\langle b_i b_j\rangle \neq 0$ higgs the internal $U(1)$ gauge field down to $Z_2$.  As argued in Ref.~\onlinecite{nandkishore2012orthogonal}, in this case the physical charge is carried by fermion $f$. In another way, we can use Ioffe-Larkin rule for physical resistivity \cite{ioffe1989gapless}: $\rho_c=\rho_b+\rho_f$. Because the boson $b$ is in a paired superfluid phase, $\rho_b=0$, therefore $\rho_c=\rho_f$. We conclude that the transport property of this phase is exactly the same as a Fermi liquid with small Fermi surfaces. Obviously we will also expect Hall number $\eta_H=-x$ as in the featureless pseudogap metal. The thermodynamic property, like specific heat or spin susceptibility should still be the same as the featureless pseudogap metal. Therefore we still view this phase as a "pseudogap metal" because the number of carriers is much smaller than the conventional Fermi liquid.

Next we will show that this pseudogap metal is a non-Fermi-liquid (NFL) instead of a Fermi liquid in terms of single electron Green function. Following the same analysis as in the previous subsection, we can get $G^c(x,y;t,t')\propto \langle b^\dagger(x,t)b(y,t') \rangle G^f(x,y;t,t')$, where we have suppressed the flavor index for simplicity. Because the schwinger boson $b$ is in a paired-superfluid phase, single  particle green function $\langle b^\dagger(x,t)b(y,t') \rangle$ should be exponential decay. As a result, single green function for the physical electron $G^c(x,y;t,t')$ should also exponential decay. As a consequence, ARPES or STM measurement can not detect any coherent quasi-particle for this pseudogap metal. The charge carrier in this exotic metal is not the physical electron. We will follow the notation of Ref.~\onlinecite{nandkishore2012orthogonal} and dub it as "orthogonal metal".

\hfill

In summary, we have shown that a featureless pseudogap metal or an orthogonal metal with small Fermi surfaces $A_{FS}=-\frac{x}{4}$ can naturally emerge at filling $\nu_T=2-x$ for large positive $J_H$ or small $J_H$.  For a negative and large $J_H$, we know that the $\nu_T=2$ Mott insulator is in the $120^\circ$ Neel order for the low energy spin-one model. Doping such a Neel order may show new metallic or superconducting phase beyond the analysis here, which we leave to future work.

\section{Deconfined metal between pseudogap metal and conventional Fermi liquid }

At the large $U>>t$ limit, we have argued that a pseudogap metal with small Fermi surfaces is likely at small doping away from $\nu_T=2$. Then  a natural question is how this small Fermi surface pseudogap metal evolves to the conventional Fermi liquid with large Fermi surfaces when increasing $\frac{t}{U}$ or the doping $x$. We try to provide one possible scenario in this section. 

For simplicity we work in the  case with $J_H=0$. We have already presented a description of the orthogonal metal with small Fermi surfaces. Next we need to understand how to describe the conventional Fermi liquid with large Fermi surfaces. At large $\frac{t}{U}$, this is a trivial problem. At the large $U>>t$ with large doping $x$, we still expect a conventional Fermi liquid, which can not be simply understood as in the free fermion case because of the constraint $P_2$ in Eq.~\ref{eq:unconventional_t_J}. In the familiar spin $1/2$ case or in the filling $\nu_T=1-x$ of the $U(4)$ model, the constraint in Eq.~\ref{eq:conventional_t_J} can be respected in the slave boson description: $c_\alpha(\mathbf x)=b^\dagger(\mathbf x) f_\alpha(\mathbf x)$. Then the conventional Fermi liquid just corresponds to the ansatz with $\langle b(\mathbf x) \rangle \neq 0 $.  In contrast, in the case $\nu_T=2-x$, the Hilbert space consists of singlon and doublon. Neither of them can simply condense without breaking spin rotation symmetry. Meanwhile, the density of singlon is $x$ while the number of doublon is $1-x$. Therefore to have the conventional Fermi liquid with Hall number $\eta_H=2-x$, we need both singlons and doublons to be absorbed to form the large Fermi surface with area $A_{FS}=\frac{2-x}{4}$. In the following we will show that a large Fermi surface state can be generated from "Kondo resonance" similar to heavy fermion systems.

To impose the constraint $P_2$ in Eq.~\ref{eq:unconventional_t_J}, we still use the parton construction in Eq.~\ref{eq:six_flavor_parton_electron}.  With this parton construction, we can define spin operators for the doublon site using the slave boson $b_a$ (linear transformation of $\Phi_{\alpha \beta}$) and the spin operators for the singlon site using the fermion $f$.   

The spin operator for the doublon site is 
\begin{equation}
 	S^{ab}_b(\mathbf x)=2i (b^\dagger_a(\mathbf x)b_b(\mathbf x)-b^\dagger_b(\mathbf x)b_a(\mathbf x))
 \end{equation} 
 where $a,b=1,2,...,6$.  $S^{ab}$ is the generator of the $SO(6)$ group. It has a one to one correspondence to the $SU(4)$ spin operators as defined in Table.~\ref{table:so(6)_generators}.

 The spin operator for the singlon site is:
 \begin{equation}
 	S^{\mu \nu}_f(\mathbf x)= f^\dagger_\alpha \tau^\mu \sigma^\nu f_\beta(\mathbf x)
 \end{equation}
 where $\alpha,\beta=1,2,3,4$.  $\tau^\mu\sigma^\nu$ with $\mu,\nu=0,x,y,z$ except $\mu=\nu=0$. 

 With this six-flavor slave boson parton construction, the $t$ term in the $t-J$ model defined in Eq.~\ref{eq:unconventional_t_J} looks like an exchange term between the singlon $f$ and the doublon $b$ : $b^\dagger_i b_j f_i f^\dagger_j$. Here we suppressed the flavor index because generically it looks quite complicated and involves many different terms.  This term can be decoupled to the mean field ansatz in Eq.~\ref{eq:six_flavor_meanfield}.  The $J$ term in this case involves terms like $S_{b;i}S_{b;j}$, $S_{f;i}S_{f;j}$ and $S_{b;i}S_{f;j}$.  At small doping $x$, we have $n_b=1-x$ and $n_f=x$. In this case we can view the doublon site as a $SO(6)$ spin and the fermion $f$ couples to the $SO(6)$ moment through the term $S_{b;i}S_{f;j}$, which resembles a kondo coupling in the heavy fermion problem.  Then a large Fermi surface may be generated through "Kondo resonance".

 The essential point of "Kondo resonance" is to absorb the $SO(6)$ spin to form a large Fermi surface with area $A_{FS}=\frac{2-x}{4}$.  To do that, we need to split a doublon to two Fermions first.  Therefore we do a further parton construction on top of Eq.~\ref{eq:six_flavor_parton_electron}:
 \begin{equation}
 	\Phi_{\alpha\beta}=\frac{1}{2}(\psi_\alpha(\mathbf x)\psi_\beta(\mathbf x)-\psi_\beta(\mathbf x)\psi_\alpha(\mathbf x))
 \end{equation}

 In another word, the original electron operator is now written:
 \begin{equation}
 	c_{\alpha}(\mathbf x)=\sum_{\beta \neq \alpha} \psi_{\alpha}(\mathbf x) \psi_{\beta}(\mathbf x) f^\dagger_{\beta}(\mathbf x)
 	\label{eq:three_parton}
 \end{equation}
 with the constraint $\frac{1}{2}n_{\psi}(\mathbf x)+n_f(\mathbf x)=1$. 

 Still there is a $U(1)$ gauge redundancy
 \begin{equation}
 	f_{\alpha}(\mathbf x)\rightarrow f_{\alpha}(\mathbf x)e^{i\varphi(\mathbf x)}\ \ \ \psi_\alpha(\mathbf x)\rightarrow e^{i\frac{1}{2}\varphi(\mathbf x)}
 \end{equation}

 So $\psi$ couples to the $U(1)$ gauge field $\mathbf a$ with gauge charge $1/2$. In addition, $\psi$ couples to another $Z_2$ gauge field because $\psi_\alpha(\mathbf x)\rightarrow -\psi_\alpha(\mathbf x)$ also does not change the physical operator.

\begin{figure}
\centering
\includegraphics[width=0.45\textwidth]{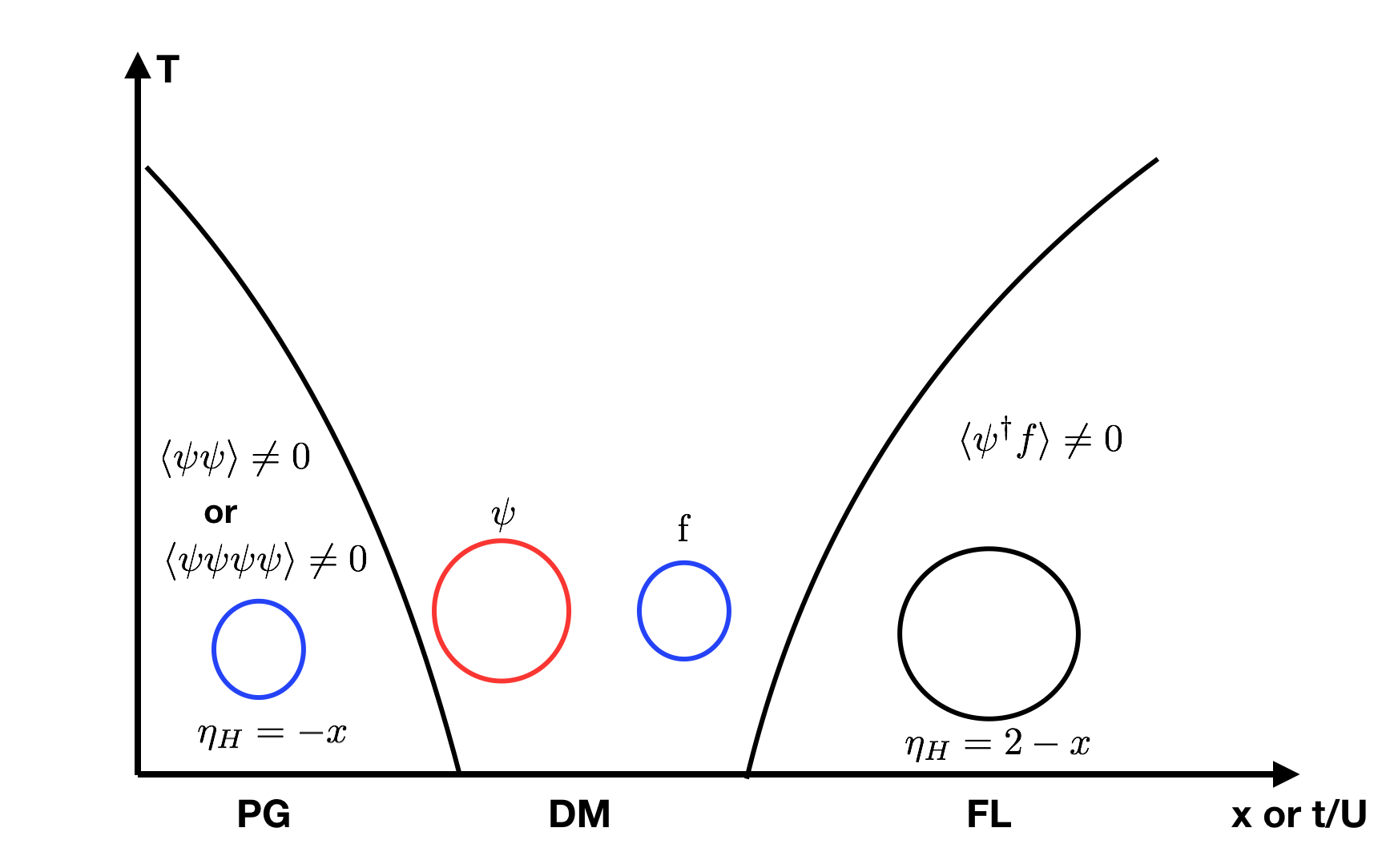}
\caption{Evolution from pseudogap metal to the conventional Fermi liquid. The red, blue and black circles denote Fermi surfaces with area $\frac{2-2x}{4}$, $\frac{x}{4}$ and $\frac{2-x}{4}$. The intermediate phase is a "deconfined metal" where the fermi surfaces formed by $\psi$ and $f$ couple to an internal $U(1)$ gauge field $\mathbf a$.  We can condense either $\psi \psi V$ or $\psi \psi \psi \psi V$ to get featureless pseudogap metal or orthogonal pseudogap metal, where $V$ is the vison annihilation operator.}
\label{DM_phase_diagram}
\end{figure}

With the parton construction in Eq.~\ref{eq:three_parton}, we can access both the conventional Fermi liquid and the pseudogap metals with small Fermi surfaces. For simplicity, let us assume that the physical gauge field $A$ couples to $f$. It turns out that the conventional Fermi liquid can evove to the pseudogap metal through two continuous transitions with an intermediate "deconfined metal", as is illustrated in Fig.~\ref{DM_phase_diagram}.  We have density $n_\psi=2-2x$ and $n_f=x$. The conventional Fermi liquid is described by $\langle \psi^\dagger f \rangle \neq 0$, similar to the "Kondo resonance" in the heavy fermion problems.  To get the pseudogap metal, we can just gap out the Fermi surfaces formed by $\psi$ through pairing.  The first possibility is a charge $2e$ pairing $\langle \psi \psi \rangle \neq 0$.  In the case with $J_H=0$, this pairing term needs to break the $SU(4)$ spin rotation and lives in a manifold generated by $SO(6)$ rotation. If $J_H>0$, we can just make $\psi$ to form the inter-valley, spin singlet pairing which preserves spin rotation symmetry.  The pairing term will completely higgs the $U(1)$ gauge field $\mathbf{a}$. However, the $Z_2$ gauge field still survives and decouples with the remaining fermi surfaces formed by $f$. This is actually a FL* phase with Fermi liquid coexisted with $Z_2$ gauge field. To avoid such an exotic state and get the featureless pseudogap metal, we should condense $\langle \psi \psi V\rangle \neq 0$ where $V$ is the vison of the $Z_2$ gauge field. In this way the $Z_2$ gauge field also gets confined and we get exactly the featureless pseudogap metal.  In the $SU(4)$ symmetric point, the orthogonal pseudogap metal may be favored. We can reach it through a charge $4e$ $SU(4)$ singlet pairing: $\langle  \epsilon_{\alpha \beta \gamma \delta} \psi_\alpha \psi_\beta \psi_\gamma \psi_\delta V\rangle \neq 0$. This pairing higgses the $U(1)$ gauge field $\mathbf{a}$ down to $Z_2$. Therefore the fermi surface from $f$ couples to both $A$ and the $Z_2$ gauge field, which is exactly the property of the orthogonal pseudogap metal described in the previous section.

Now we have one theory that the pseudogap metal can evolve to the Fermi liquid through two continuous phase transitions. Next we briefly discuss the property of the intermediate deconfined metal.  The low energy physics is governed by the following action:
\begin{align}
	L&=\psi^\dagger (\partial_0-\frac{1}{2}a_0)\psi +f^\dagger (\partial_0-a_0)f \notag\\
	&+\frac{\hbar^2}{2m_\psi}\psi^\dagger(-i\vec \partial-\frac{1}{2}\vec a)^2\psi+\frac{\hbar^2}{2m_f}f^\dagger(-i\vec \partial+\vec A-\vec a)^2f
\end{align}

In gaussian approximation, an Ioffe-Larkin rule can be easily derived as:
\begin{equation}
	\rho_c=\rho_{\psi}+\rho_f
\end{equation}
where $\rho_\psi$ and $\rho_f$ should be viewed as the resistivity tensor for $\psi$ and $f$.

Because of the gauge fluctuation, the quasiparticle picture is known to break down. Transport of such a deconfined metal remains to be an unsolved theoretical problem. We hope that the possible realization of this phase in graphene moir\'e superlattice can provide more information from experiment on this problem.

When applying an external magnetic field, we can get an effective action for the internal magnetic flux:
\begin{equation}
	L_{eff}=\chi_\psi |\frac{1}{2}b|^2+\chi_f |b-B|^2
\end{equation}
In the saddle point, the internal gauge field flux $\mathbf{b}$ is locked to the external magnetic field: $\mathbf{b}=\alpha \mathbf{B}$ with $\alpha= \frac{4 \chi_f}{\chi_\psi+4\chi_f}$. Here $\chi_{\psi}$ and $\chi_{f}$ are the diamagnetism susceptibility in the gaussian approximation. Generically we expect $\alpha$ is an order one number smaller than one. Using $\chi\sim \frac{1}{m}$ for a Fermi surface, we have $\alpha=\frac{4 m_{\psi}}{4 m_{\psi}+m_f}$. Because of the locking, $\psi$ sees an effective field $\alpha \mathbf{B}$ while $f$ sees an effective field $-(1-\alpha)\mathbf{B}$. Then in principle one should see quantum oscillations corresponding to the Fermi surfaces area for both $\psi$ and $f$ with frequency renormalized by factor $\alpha$ and $1-\alpha$. 

Next we discuss the Hall number of this deconfined metal. The constraint $\frac{1}{2}n_\psi+n_f=1$ implies that $\mathbf{J}_\psi=\mathbf{J}_f$, where $\mathbf{J}_\psi=\mathbf{J}$ is defined as variation of $\mathbf{a}$ while $\mathbf{J}_f$ is defined as variation of $\mathbf{A}$. In gaussian approximation, we have the result:
\begin{align}
e_x&=\frac{\alpha B}{n_\psi \frac{1}{2}e}J^\psi_y\notag\\
e_x-E_x&=\frac{1-\alpha}{n_fe}B J^f_y
\end{align}
where we have used the result $\rho_{xy}=\frac{B}{n Q}$ for a Fermi surface with charge $Q$ and density $n$.

Then it is easy to get
\begin{equation}
	E_x=\left( \frac{2\alpha}{n_\psi e}-\frac{1-\alpha}{n_f e} \right) B J_y
\end{equation}

Therefore the Hall number is
\begin{equation}
	\frac{1}{\eta_H}=(\frac{2\alpha}{n_\psi }-\frac{1-\alpha}{n_f})
\end{equation}
or
\begin{equation}
	\eta_H=\left(\frac{\alpha}{1-x}-\frac{1-\alpha}{x}\right)^{-1}
\end{equation}

Note that generically $\eta_H$ is not related to the density of charge carriers. At the limit $x<<1$ and $\alpha<<1$, we have $\eta_H \approx - \frac{x}{1-\alpha}$. $\eta_H$ diverges when $\alpha$ increases to $1-x$. Therefore the Hall number can be arbitrary inside the deconfined metal, depending on the value of $\alpha$.

Once the system is in the FL phase, $\langle \psi^\dagger f \rangle$ locks $\mathbf{a}=2\mathbf{A}$. In this case $\psi$ forms an electron pocket while $f$ can be viewed as a hole pocket. Deep inside the FL phase, the two Fermi surfaces merge together to form a large electron Fermi surface and $\eta_H=2-x$.

In the above we have described one scenario for evolution from pseudogap metals to conventional Fermi liquid by increasing doping or bandwidth. In this simple scenario there is an intermediate deconfined metal phase. The physics of the pseudogap metal we described is kind of similar to the "symmetric mass generation" proposed in Ref.~\onlinecite{you2018symmetric} for Dirac fermions. In that simple case Ref.~\onlinecite{you2018symmetric} constructed a deconfined critical point between an insulator and a Dirac semimetal. It is not clear whether a  direct transition between the pseudogap metal and the conventional Fermi liquid can exist or not in our case. We hope to study this in future.

\section{$U(1)$ Spinon Fermi surface state and $Z_4$ spin liquid at $\nu_T=1$}
At $\nu_T=1$, the inter-valley Hund's term $J_H$ vanishes after projection to the Hilbert space without double occupancy. Therefore the spin rotation symmetry is $SU(4)/Z_4$ or $U(1)_{valley}\times SO(3)_+\times SO(3)_-$.  We will show that there is no symmetric gapped $Z_2$ spin liquid when spin rotation symmetry is both $SU(4)/Z_4$ and $U(1)_{valley}\times SO(3)_+\times SO(3)_-$. Within the Abrikosov fermion parton construction, the only symmetric spin liquid state is a $U(1)$ spin liquid with spinon fermi surface, which can be reached from the Fermi liquid phase through a continuous transition. A symmetric $Z_4$  spin liquid is also possible but beyond mean field description.

\subsection{Absence of Gapped Symmetric $Z_2$ Spin Liquid}
There is a general argument to rule out  gapped $Z_2$ spin liquid with full spin rotation symmetry. Because there is only one fundamental representation within each unit cell,  in a spin rotation invariant $Z_2$ spin liquid, spinon needs to transform as $(\frac{1}{2},0)$ and $(0,\frac{1}{2})$ under $SU(2)_+ \times SU(2)_-$. Basically we have $e_{+\sigma}$ carries valley $+$ and $e_{-\sigma}$ carries valley $-$.  A bound state formed by $e_{+\sigma_1}e_{-\sigma_2}$  has dimension $4$ and transforms as $(\frac{1}{2},\frac{1}{2})$ under $SU(2)_+\times SU(2)_-$. Under $2\pi$ rotation of $S_z$ for either valley, it acquires a global $-1$ phase. Therefore the bound state is in the projective representation of $U(1)_{valley}\times SO(3)_+ \times SO(3)_-$.  However, in a gapped $Z_2$ spin liquid, $e_+ e_-$ is created by a local operator and should be in a linear representation. The contradiction implies that a gapped symmetric $Z_2$ spin liquid is not possible. 

\subsection{$U(1)$ Spin Liquid in Abrikosov Fermion Parton Construction}
A $U(1)$ spin liquid is possible and can be constructed in the  standard Abrikosov fermion parton theory. For the current problem, Schwinger boson parton theory is not very useful because there is no symmetric paired condensation phase of the four-flavor Schwinger bosons. Therefore we restrict to Abrikosov fermion parton theory.   At each site, we have $f_{i a}$ in the fundamental representation of $SU(4)$ with the constraint $\sum_a f^\dagger_{ia}f_{ia}=1$.  Unlike the spin $1/2$ case, there is only a $U(1)$ gauge redundancy.

Apparently there can not be pairing term which preserves the spin rotation symmetry. This is another manifestation that symmetric $Z_2$ spin liquid is not possible. In the following we classify all possible symmetric $U(1)$ spin liquid states.

Projective Symmetric Group (PSG) for $U(1)$ spin liquid is classified in Appendix.~\ref{appendix: U(1) PSG}. There are only two possible PSG, which is labeled by the projective translation symmetry: $T_1 T_2=T_2 T_1 e^{i \Phi_T}$.   It turns out only $\Phi_T=0$ and $\Phi_T=\pi$ are compatible with the reflection symmetry $\sigma$. Once $\Phi_T$ is fixed, the symmetry realizations of $\sigma$ and $C_6$  are also fixed. Note that $C_6^6$ and $\sigma^2$ are meaningless in a $U(1)$ spin liquid because a global $U(1)$ transformation can always be added in $C_6$ and $\sigma$.   For each symmetry operation $X$, the symmetry realization is $e^{i\varphi_X(x,y)}X$. The following is a list of the PSG.

\begin{align}
&\varphi_{T_1}=0\notag\\
&\varphi_{T_2}=p_1 \pi x \notag\\
&\varphi_{\sigma}=p_1 \pi xy\notag\\
&\varphi_{C_6}=p_1 \pi xy + \frac{1}{2}p_1 \pi  y(y-1)
\end{align}
up to a constant phase.  $\Phi_T=p_1 \pi$ with $p_1=0,1$.

$p_1=0$ and $p_1=1$ label the zero-flux phase and the $\pi$-flux phase. However, in the $\pi$ flux phase the nearest neighbor hopping and the next nearest neighbor hopping are forbidden. This ansatz is not energetically favorable. Therefore we only consider the zero flux phase.

In the zero flux phase, $\varphi_{X}=0$. Therefore all of symmetry operations are realized trivially.  The phase is a $U(1)$ spin liquid with four fermi surfaces, each at filling $1/4$. This spin liquid phase can be reached from the Fermi liquid side through a continuous quantum phase transition. In principle a symmetric $Z_4$ spin liquid is also possible. In the Abrikosov fermion description, the fermion can form a charge $4e$ singlet pairing: $\langle \epsilon_{abcd}f_af_bf_cf_d \rangle \neq 0$, resulting in a symmetric $Z_4$ spin liquid.

For the spin-valley model at $\nu_T=1$, there is no symmetric featureless Mott insulator and magnetic order may be suppressed because of the frustration of triangular lattice and large quantum fluctuation space. The most likely competing ordered state is a plaquette order with four sites forming a $SU(4)$ singlet. If such a plaquette order is melted, we can get a quantum spin liquid phase. In this case, our PSG analysis suggests a $U(1)$ spin liquid with spinon Fermi surface or a $Z_4$ spin liquid.  

A superconductor has been reported at small doping away from the $\nu_T=1$ Mott insulator\cite{Wang2019Signatures}. For both the $Z_4$ spin liquid and the plaquette order, the Mott insulator is a $SU(4)$ singlet. Therefore upon doping, the most likely superconductor has charge $4e$ pairing. A charge $4e$ superconductor will be killed by in plane Zeeman field, which is consistent with the experiment\cite{Wang2019Signatures}. In contrast, a conventional charge $2e$ pairing lives on a $SO(4)$ manifold because valley triplet, spin singlet pairing is degenerate with the valley singlet, spin triplet pairing. A zeeman field will select the spin triplet pairing and there is no reason to expect the $T_c$ to be suppressed by Zeeman field.  Given the experimental phenomenology and the possible $SU(4)$ symmetric Mott insulator nearby, the possibility that the observed superconductor is charge $4e$ paired should be taken seriously. We leave a detailed analysis of charge $4e$ superconductor to future work.

\section{Conclusion}
In this paper we study possible interesting phases in a spin-valley Hubbard model on triangular moire superlattice. We show that pseudogap metals with small Fermi surfaces can naturally emerge by doping the $\nu_T=2$ Mott insulator. In the  moir\'e materials, it is also easy to study the possible transition between the "pseudogap metals" and the conventional Fermi liquid by tuning either doping or displacement field. We propose one possible route through an intermediate deconfined metallic phase.  We also comment on possible spin liquids at $\nu_T=1$ and charge $4e$ superconductor nearby. Our proposals can be easily tested in ABC trilayer graphene aligned with hBN and in twisted transition metal dichalcogenide homobilayers.  The existence of two distinct symmetric Fermi liquids from symmetry analysis is also true for the graphene moir\'e systems with topological bands. In future it is interesting to study whether a similar symmetric Fermi liquid with small Fermi surfaces can naturally exist in the topological case. 

\section{Acknowledgement}

We thank T. Senthil for very helpful discussions. This work was supported by US Department of Energy grant DE- SC000873

\bibliographystyle{apsrev4-1}
\bibliography{spin_liquid}

\onecolumngrid
\appendix

\section{Crystal Symmetry of Triangular Lattice\label{append:lattice_symmetries}}
We define $\mathbf{r}=x \mathbf{a_1}+\mathbf{a_2}$. The lattice symmetries are:
\begin{align}
&T_1: (x,y)\rightarrow(x+1,y)\notag\\
& T_2: (x,y)\rightarrow(x,y+1)\notag\\
&\sigma: (x,y)\rightarrow (y,x)\notag\\
&C_6: (x,y)\rightarrow(x-y,x)
\end{align}

The following algebraic constraints are useful.
\begin{align}
&T_2 T_1=T_1 T_2 \notag\\
&T_1 C_6=C_6 T_2^{-1}\notag\\
&T_2 C_6=C_6 T_1 T_2 \notag\\
&T_1 \sigma=\sigma T_2 \notag\\
&T_2 \sigma=\sigma T_1 \notag\\
&C_6^6=1\notag\\
&\sigma^2=1 \notag\\
&C_6 \sigma=\sigma C_6^5
\end{align}

\section{Projective Symmetry Group Classification for $U(1)$ Spin Liquid\label{appendix: U(1) PSG}}
At $\nu_T=1$, we can use the Abrikosov fermion in the $SU(4)$ fundamental representation. At $\nu_T=2$, we can have $U(1)$ spin liquid described by six flavor Schwinger fermion. In this section we classify all symmetric $U(1)$ spin liquid states within the  fermion parton theory for both $\nu_T=1$ and $\nu_T=2$.

First, IGG is $\{ e^{i \theta}\}$ where $\theta \in [0,2\pi)$ is a constant phase.   For each symmetry operation $X$, we can parametrize the gauge transformation as $G_{X}(\mathbf r)=e^{i \varphi_{\mathbf X}(\mathbf r)}$.

Under gauge transformation $G(\mathbf x)=e^{i\varphi_G(\mathbf x)}$, the $G_X$ should be replaced by $G G_X X G^{-1}X^{-1}$. Correspondingly we have:

\begin{align}
\varphi_X(\mathbf r)\rightarrow \varphi_G(\mathbf r)+\varphi_X(\mathbf r)- \varphi_G(X^{-1} \mathbf r)
\end{align}

Next we need to fix the gauge. Following Wen and  Fa Wang et.al \cite{wang2006spin}, we fix  $\varphi_{T_1}(\mathbf r)=0$. This can be done by solving the equations:
\begin{align}
\varphi_G(x,y)+\varphi_X(x,y)- \varphi_G(x-1,y)=0
\label{eq:fix_T_1_phase}
\end{align}

These two equations fix dependence of $G(x,y)$ on $x$.  If $G(0,y)$ is fixed, then $G(x,y)$ is fixed. Now we only have the gauge freedom $G(0,y)$. Then we can use $T_1 T_2=T_2 T_1$ to fix $G_{T_2}$.  Using $T_1^{-1}T_2 T_1 T_2^{-1}=I$, we have:
\begin{align}
&\varphi_{T_2}(x,y)=\varphi_{T_2}(0,y)+\Phi_T x
\end{align}
where  $\Phi \in [0,2\pi)$ is a position independent constant.

We can use  the remaining gauge freedom $G(0,y)$ to make $\varphi_{T_2}(0,y)=0$.  Because IGG is $U(1)$, a constant phase in $\varphi_X(x,y)$ does not matter. A non-zero $\Phi$ means a projective translation symmetry: $T_2T_1=T_1 T_2 e^{i \Phi_T}$. We need to fix $\frac{\Phi_T}{\pi}=\frac{p}{q}$ where $p,q$ are integers.

Next we need to find PSG for $\sigma$ and $C_6$.  First we need to point out the remaining gauge freedom we can use. The first one is $G_1: \varphi_1=constant$.  The second one is $G_2: \varphi_2(x,y)=\theta_2 x$.  This will change $\varphi_{T_1}=\theta_2$. However, it belongs to the IGG and does not matter. The third one is $G_3: \varphi_3(x,y)=\theta_3 y$. These gauge of freedom can be used to eliminate redundant parameters later.

Finally we get $\varphi_\sigma(x,y)=\Phi_T xy+Constant$. From $T_1^{-1}\sigma T_2 \sigma^{-1}=I$ and $T_2^{-1}\sigma T_1 \sigma^{-1}=I$ we have:
\begin{align}
& \varphi_{\sigma}(x+1,y)-\varphi_{\sigma}(x,y)=-\Phi_T y  +\phi'_2\notag\\
& \varphi_{\sigma}(x,y+1)-\varphi_{\sigma}(x,y)=\Phi_T x  +\phi'_3
\end{align}
where $\phi'_2,\phi'_3$ are constant phases.

From these equations we can get $\varphi_{\sigma}(X,Y)=\varphi_{\sigma}(0,0)+\phi'_2 X+\phi'_3 Y+\Phi_T XY\ mod\ 2\pi=\varphi_{\sigma}(0,0)+\phi'_2X+\phi'_3 Y-\Phi_T X Y\ mod\ 2\pi$. To have solution, we need to fix $\Phi_T=0,\pi$. Therefore a general flux $\Phi$ is not compatible with the reflection symmetry.  In the following we can use the notation $\Phi_T=p_1 \pi$. A general solution is:
\begin{equation}
	\varphi_{\sigma}(x,y)=\varphi_{\sigma}(0,0)+p_1 \pi xy+\phi'_2 x+\phi'_3 y
\end{equation}

From $\sigma^2=I$ we have:
\begin{equation}
	\varphi_{\sigma}(x,y)+\varphi_{\sigma}(y,x)=2\varphi_{\sigma}(0,0)
\end{equation}
which fixes $\phi'_2=-\phi'_3\ mod\ 2\pi$.

Now we can use the gauge freedome $G_2=e^{i \theta_2 x}$ to reduce the parameters. $\varphi_{\sigma}(x,y)$ changes to
\begin{align}
	&\varphi_\sigma(x,y)\rightarrow  \varphi_\sigma(x,y)+\theta_2(x-y)\notag\\
	&\rightarrow\varphi_\sigma(0,0)+p_1 \pi xy+(\phi'_2+\theta_2)(x-y)
\end{align}
We can always choose $\theta_2=-\phi'_2$.  Finally we have the PSG for $\sigma$:
\begin{equation}
	\varphi_\sigma(x,y)=p_1 \pi xy+constant
\end{equation}

The next task is $C_6$. We can still use $G_1$ and $G_2G_3$, which do not change $G_{T_1},G_{T_2}, G_\sigma$ up to a constant phase.

Using $T_1^{-1} C_6 T_2^{-1}C_6^{-1}=I$ and $T_2^{-1}C_6T_1 T_2 C_6^{-1}=I$, we can get:
\begin{align}
& \varphi_{C_6}(x+1,y)-\varphi_{C_6}(x,y)=\phi'_4+p_1 \pi y\notag\\
& \varphi_{C_6}(x,y+1)-\varphi_{C_6}(x,y)=\phi'_5+p_1 \pi (x-y)
\end{align}
where $\phi'_4,\phi'_5$ are constant $U(1)$ phases.

A general solution is:
\begin{equation}
	\varphi_{C_6}(x,y)=\varphi_{C_6}(0,0)+p_1 \pi xy-\frac{1}{2}p_1 y(y-1)\pi+\phi'_4 x+\phi'_5 y
\end{equation}

$C_6 \sigma C_6 \sigma$ further impose the constraint

\begin{equation}
	\varphi_{C_6}(x,y)+\varphi_{C_6}(y-x,y)+\varphi_{\sigma}(y,y-x)+\varphi_\sigma(y,x)=constant
\end{equation}
which fixes $\phi'_4+2 \phi'_5=0 \ mod\ 2\pi$.

$C_6^6=I$ imposes the constraint $\phi'_4=0 \ mod \ 2\pi$. Then we can also fix $\phi'_5=p'_5 \pi$ with $p'_5=0,1$.

Under the gauge transformation $G'_3=e^{i\theta_3(x+y)}$, the solution of $\varphi_{C_6}$ changes to:
\begin{equation}
	\varphi_{C_6}\rightarrow \varphi_{C_6}(0,0)+p_1 \pi xy-\frac{1}{2}p_1 y(y-1)\pi+(p'_5 \pi-\theta_3) y+2\theta_3 x
\end{equation}

Choosing $\theta_3=p'_5 \pi$, $\varphi_{C_6}$ can always be reduced to:
\begin{equation}
	\varphi_{C_6} = p_1 \pi xy+\frac{1}{2}p_1 y(y-1)\pi
\end{equation}
up to a constant phase.

\section{Mean Field of $Z_2$ spin liquid at $\nu_T=2$ \label{append:Z_2_spin_liquid}}

In this section we analyze mean field ansatz for $SO(6)$ symmetric Hamiltonian with only nearest neighbor coupling $J$  at $\nu_T=2$ based on the Schwinger fermion and the Schwinger boson parton theories.We focus on the two symmetric $Z_2$ spin liquids.

\subsection{Type I $Z_2$ Spin Liquid}
The type I $Z_2$ spin liquid has PSG $P^b=(0,0,0)$ and $P^f=(1,1,0)$. The bosonic $e$ spinon has a trivial PSG while the $f$ particle is in a $\pi$ flux phase. We can get the dispersions of the $e$ particle and the $f$ particle from the Schwinger boson and the Schwinger fermion mean field theories respectively.

\subsubsection{Mean Field theory for the Bosonic Spinon}
The dispersion of the $e$ particle is described by the Schwinger boson mean field theory with zero-flux ansatz. We have both nearest neighbor hopping and pairing terms.

Because all of the six bosons decouple with each other in the mean field level,  we can work with a spinless boson $b$ at filling $n_b=\frac{1}{6}$ with the Hamiltonian:
\begin{equation}
	H_b=t\sum_{\langle ij \rangle}b^\dagger_i b_j-\Delta^* \sum_{\langle ij \rangle}(b_i b_j+h.c.)-\mu \sum_{i}b^\dagger_i b_i
\end{equation}

or in momentum space:
\begin{align}
	H_b&=\frac{1}{2} \sum_{\mathbf k} (b^\dagger(\mathbf k),b(-\mathbf k))
	\left(
	\begin{array}{cc}
	\xi(\mathbf k) & -\Delta(\mathbf k)\\
	-\Delta^*(\mathbf k) & \xi(\mathbf k)
	\end{array}
	\right)
	\left(
	\begin{array}{c}
	b(\mathbf k)\\
	b^\dagger(-\mathbf k)
	\end{array}
	\right)\notag\\
	&+...
\end{align}
where $\Delta(\mathbf k)=\sum_{\mathbf R}\Delta e^{-i \mathbf{k}\cdot \mathbf {R}}$

Using the standard Bogoliubov transformation:
\begin{equation}
	\alpha_{\mathbf k}=\mu_k b_k+\upsilon_k b^\dagger_{-k}
\end{equation}
with the constraint:
\begin{equation}
	\mu^2(\mathbf k)-\upsilon^2(\mathbf k)=1
\end{equation}

The inverse transformation is
\begin{equation}
	b_{\mathbf k}=\mu^*_{\mathbf k}\alpha_{\mathbf k}-\upsilon_{\mathbf k} \alpha^\dagger_{-\mathbf k}
\end{equation}
where we assumed $\mu_{\mathbf k}=\mu_{-\mathbf k}$ and $\upsilon_{\mathbf k}=\upsilon_{-\mathbf k}$.

The solution is
\begin{align}
	\mu^2_{\mathbf k}&=\frac{1}{2}\left(\frac{\xi(\mathbf k)}{E_{\mathbf k}}+1\right)\notag\\
	\upsilon^2_{\mathbf k}&=\frac{1}{2}\left(\frac{\xi(\mathbf k)}{E_{\mathbf k}}-1\right)
\end{align}
with the sign
\begin{equation}
	2 \mu_{\mathbf k}\upsilon_{\mathbf k}=-\frac{\Delta}{E_{\mathbf k}}
\end{equation}

where,
\begin{equation}
	E_{\mathbf k}=\sqrt{\xi_{\mathbf k}^2-\Delta^2_{\mathbf k}}
\end{equation}

The final dispersion is:
\begin{equation}
	H=\frac{1}{2} \sum_{\mathbf k}E_{\mathbf k}(\alpha^\dagger_{\mathbf k}\alpha_{\mathbf k}+\alpha^\dagger_{-\mathbf k}\alpha_{-\mathbf k})+E_0
\end{equation}

At zero $T$, the expectation value is $\langle \alpha^\dagger_{\mathbf k}\alpha_{\mathbf k} \rangle =0$, which leads to
\begin{equation}
	\langle b^\dagger_{\mathbf k}b_{\mathbf k}\rangle=\upsilon^2_{\mathbf k}
\end{equation}
and
\begin{equation}
	\langle b_{\mathbf k}b_{-\mathbf k} \rangle=-\mu_{\mathbf k}\upsilon_{\mathbf k}=\frac{\Delta_{\mathbf k}}{2 E_{\mathbf k}}
\end{equation}

Finally, we get the self consistent equaltion:
\begin{align}
&\frac{1}{N}\sum_{\mathbf k}\upsilon^2_{\mathbf k}=\frac{1}{6}\notag\\
&t=24(1+\frac{1}{6})J \frac{1}{N}\sum_{\mathbf k}\upsilon^2_{\mathbf k}e^{i \mathbf{k}\cdot \mathbf{a_1}}\notag\\
&\Delta=24(1+\frac{1}{6})J \frac{1}{N}\sum_{\mathbf k}\frac{\Delta_{\mathbf k}}{2 E_{\mathbf k}}e^{-i \mathbf{k}\cdot \mathbf{a_1}}
\end{align}
where $\mathbf{a_1}=(1,0)$ is one bond.

The mean field energy is:
\begin{align}
	E_M&=28J \sum_{\langle ij \rangle} (\langle b_i^\dagger b_j \rangle^2-\langle b_i b_j \rangle^2) \notag\\
\end{align}

For  total density $n_b=0.5/6$, we find ansatz with $t=0.795 J$ and $\Delta=3.644 J$. Such an ansatz has band bottom at the $\Gamma$ point. The mean field energy is $E_M=- 0.03732 \times 42 J$.   The critical density for condensation is around $1.45$.

The condensation of $e$ particle leads to a Ferromagnetic state, which is apparently not physical for an anti-ferromagnetic spin model.

\subsubsection{Mean field theory of $f$ particle}
\begin{figure}[H]
\centering
\includegraphics[width=0.5\textwidth]{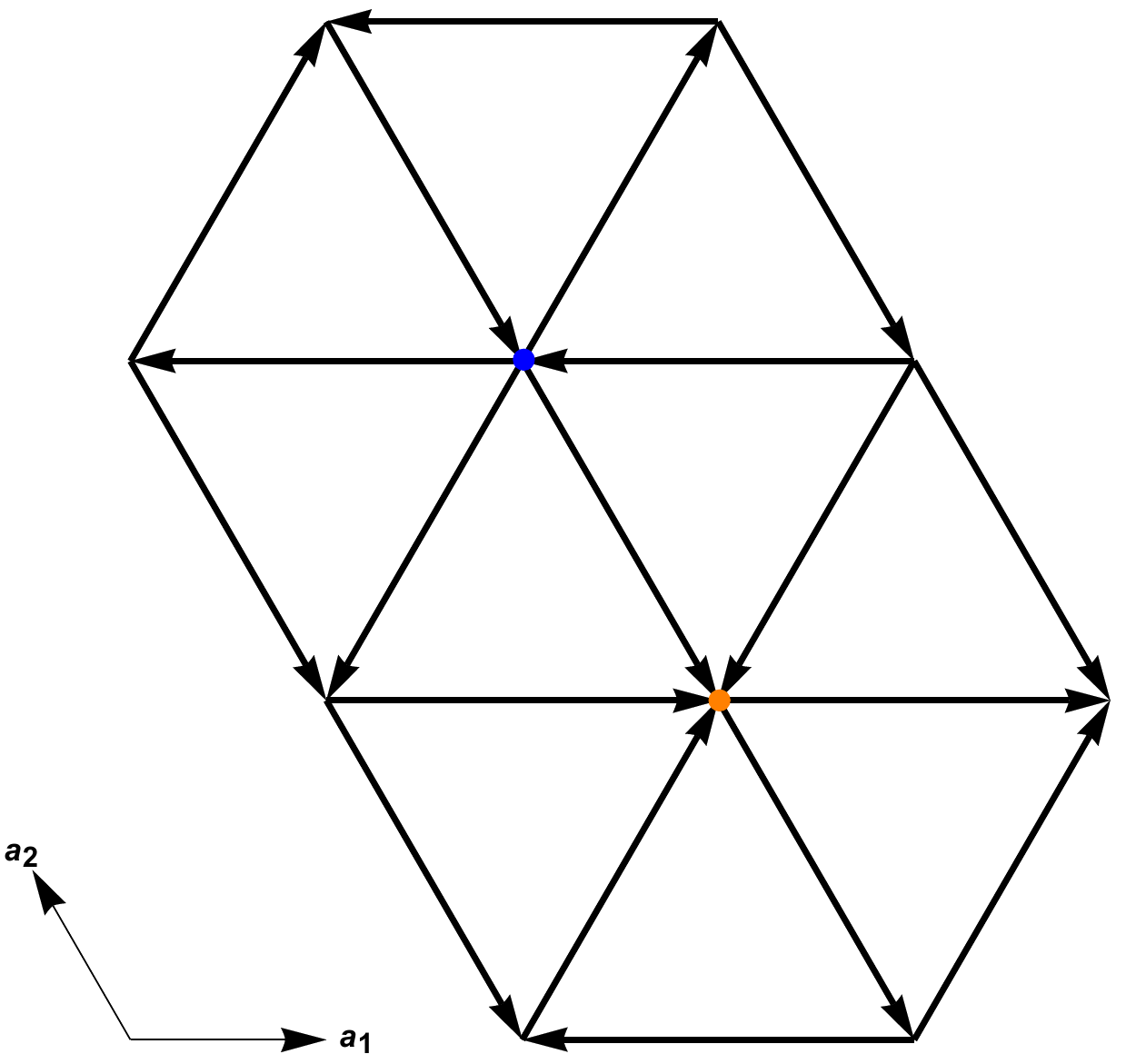}
\caption{$\pi$ flux ansatz for Schwinger fermion with nearest neighbor pairing. The pairing term is odd under inversion. The direction of the arrow denotes the positive pairing.}
\label{fig:fermion_pi_flux}
\end{figure}

Schwinger fermion mean field theory describes the dispersion of the $f$ particle.  The PSG is $P^f=(1,1,0)$. It has zero nearest neighbor hopping while the pairing terms follow a $\pi$ flux ansatz depicted in Fig.~\ref{fig:fermion_pi_flux}.

 The mean field ansatz  for each  component is effectively spinless.

\begin{equation}
	H_M=-\mu \sum_{i}f_i^\dagger f_i- \sum_{\langle ij \rangle}(\Delta_{ij}^*f_i f_j+h.c.)
\end{equation}

We need to fix the filling $\langle f_i^\dagger f_i \rangle =\frac{1}{6}$. The pairing ansatz is $\Delta_{ij}=-\Delta_{ji}$ with $|\Delta_{ij}|=\Delta$.

The $\pi$ flux ansatz has two sublattices $A$ and $B$. In the momentum space,
\begin{equation}
	H=\frac{1}{2}\sum_{\mathbf k \in sBZ}(f^\dagger_A(\mathbf k),f^\dagger_B(\mathbf k),f_A(-\mathbf k), f_B(-\mathbf k))H(\mathbf k)
	\left(\begin{array}{c}f_{A}(\mathbf k)\\f_B(\mathbf k)\\f^\dagger_A(-\mathbf k)\\f^\dagger_B(-\mathbf k)\end{array}\right)
\end{equation}
where $sBZ$ is a smaller Rectangular BZ with half area. $H(\mathbf k)$ is
\begin{equation}
	H(\mathbf k)=\left(
	\begin{array}{cc}
	-\mu I & P(\mathbf k)^\dagger\\
	P(\mathbf k) & \mu I
	\end{array}
	\right)
\end{equation}
where $I$ is the Identity matrix with $2\times 2$ dimension. $P(\mathbf k)$ is:
\begin{equation}
	P(\mathbf k)=2i\Delta (\sin k_1 \delta_z+\sin k_2 \sigma_x -\cos k_3 \sigma_y)
\end{equation}
where $k_1=\mathbf{k}\cdot \mathbf{a_2}=k_x$, $k_2=\mathbf{k}\cdot \mathbf{a_2}=-\frac{1}{2}k_x+\frac{\sqrt{3}}{2}k_y$ and $k_3=\mathbf{k}\cdot (\mathbf{a_1+a_2})=\frac{1}{2}k_x+\frac{\sqrt{3}}{2}k_y$.

The dispersion is
\begin{equation}
	E(\mathbf k)=\sqrt{\mu^2+4 \Delta^2(\sin^2 k_1+\sin^2 k_2+\cos ^2 k_3)}
\end{equation}

Self consistently we find $\Delta=4.213 J$ with mean field energy $E_M=-0.0226 \times 42 J$. $\mu=-8.93 J$.

\subsection{Type II $Z_2$ Spin Liquid}

The type II $Z_2$ spin liquid has PSG $P^b=(1,1,1)$ and $P^f=(0,0,1)$. The bosonic $e$ spinon is in a $\pi$ flux phase while the $f$ particle is in a zero flux phase. We can get the dispersions of the $e$ particle and the $f$ particle from the Schwinger boson and the Schwinger fermion mean field theoreis respectively.

\subsubsection{Mean Field theory of $e$ particle}

\begin{figure}[H]
\centering
\includegraphics[width=0.4\textwidth]{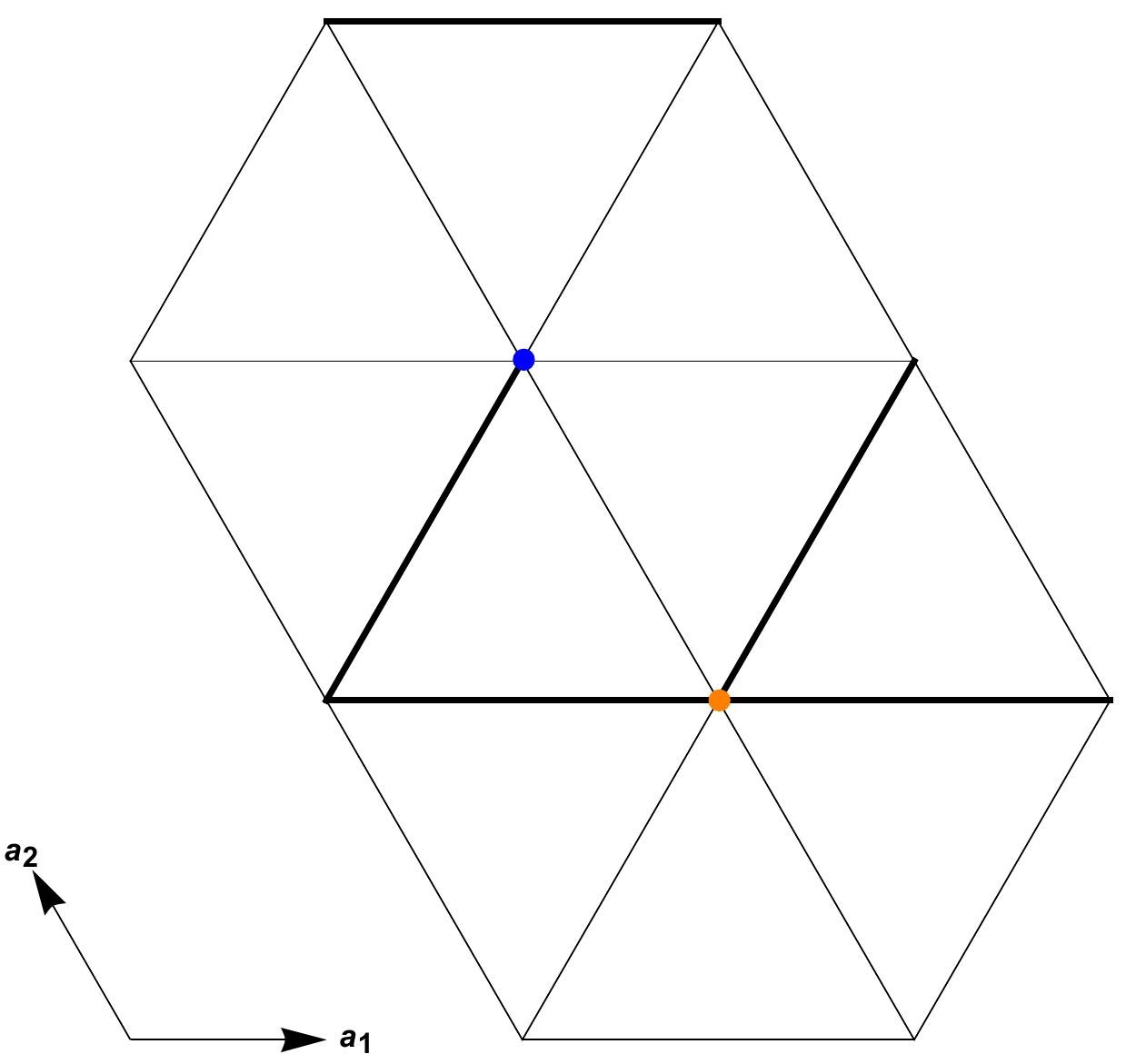}
\caption{$\pi$ flux ansatz for Schwinger boson with nearest neighbor pairing. The pairing term is even under inversion. The bold bond denotes positive pairing while the weak bond denotes negative pairing. There are two inequivalent sublattices.}
\label{fig:boson_pi_flux}
\end{figure}

In the $\pi$ flux state, the unit cell is doubled. We have two sublattices $A$ and $B$. There is only pairing term  with pattern shown in Fig.~\ref{fig:boson_pi_flux}.

  The Hamiltonian is:
\begin{equation}
	H=\frac{1}{2}\sum_{\mathbf k \in sBZ}(b^\dagger_A(\mathbf k),b^\dagger_B(\mathbf k),b_A(-\mathbf k), b_B(-\mathbf k))H(\mathbf k)
	\left(\begin{array}{c}b_{A}(\mathbf k)\\b_B(\mathbf k)\\b^\dagger_A(-\mathbf k)\\b^\dagger_B(-\mathbf k)\end{array}\right)
\end{equation}
where $sBZ$ is a smaller Rectangular BZ with half area. $H(\mathbf k)$ is
\begin{equation}
	H(\mathbf k)=\left(
	\begin{array}{cc}
	-\mu I & P(\mathbf k)\\
	P(\mathbf k)^\dagger & -\mu I
	\end{array}
	\right)
\end{equation}
where $I$ is the Identity matrix with $2\times 2$ dimension. $P(\mathbf k)$ is:
\begin{equation}
	P(\mathbf k)=2\Delta (\cos k_1 \delta_z+\cos k_2 \sigma_x+\sin k_3 \sigma_y)
\end{equation}
where $k_1=\mathbf{k}\cdot \mathbf{a_2}=k_x$, $k_2=\mathbf{k}\cdot \mathbf{a_2}=-\frac{1}{2}k_x+\frac{\sqrt{3}}{2}k_y$ and $k_3=\mathbf{k}\cdot (\mathbf{a_2-a_1})=\frac{1}{2}k_x+\frac{\sqrt{3}}{2}k_y$

The energy spectrum is
\begin{equation}
	E_{\mathbf k}=\sqrt{\mu^2-4\Delta^2(\cos^2 k_1+\cos^2 k_3+\sin^2 k_2)}
\end{equation}

The self consistent equations are:
\begin{align}
&\frac{1}{2}\sum_{\mathbf k \in sBZ}\left(\frac{|\mu|}{E_{\mathbf k}}-1\right)=\frac{1}{6}\notag\\
&28 J\sum_{\mathbf k \in sBZ}\frac{4\Delta (\cos^2 k_1+\cos^2 k_3+\sin^2 k_2)}{12 E_{\mathbf k}}=\Delta
\end{align}

The sBZ is defined as $k_x \in [-\pi, \pi] $ and $k_y \in [-\frac{\pi}{\sqrt{3}},\frac{\pi}{\sqrt{3}}]$.

The solution is $\Delta=5.1095 J$ with mean field energy $E_{MF}=-0.0333 \times 42 J$.  The critical density for condensation is around $5.09$, which is quite large.

 The band minimums of the dispersion are at the following four points: $\mathbf{K_1}=(\frac{\pi}{6},\frac{\pi}{2\sqrt{3}})$, $\mathbf{K_2}=(\frac{5\pi}{6},\frac{\pi}{2\sqrt{3}})$, $-\mathbf{K_1}$ and $-\mathbf{K_2}$. Condensation of $e$ particle can therefore lead to antiferromagnetic order. Two spinon spectrum minimum: $\mathbf{Q_1}=\mathbf{K_2}-\mathbf{K_1}=(\frac{2\pi}{3},0)$, $\mathbf{Q_2}=\mathbf{K_1}-(\mathbf{-K_1})=(\frac{\pi}{3},\frac{\pi}{\sqrt{3}})$, $\mathbf{Q_3}=\mathbf{K_2}-(-\mathbf{K_2})=(-\frac{\pi}{3},\frac{\pi}{\sqrt{3}})$, $\mathbf{Q_4}=\mathbf{K_1}-\mathbf{K_2}=(-\frac{2\pi}{3},0)$,  $\mathbf{Q_2}=(\mathbf{-K_1})-\mathbf{K_1}=(-\frac{\pi}{3},-\frac{\pi}{\sqrt{3}})$ and $\mathbf{Q_6}=(-\mathbf{K_2})-\mathbf{K_2}=(\frac{\pi}{3}.-\frac{\pi}{\sqrt{3}})$.  These six vectors form another Hexagon.  The enlarged unit cell is $2\sqrt{3}\times 2\sqrt{3}$.

\subsubsection{Mean field theory of $f$ particle}
 
\begin{figure}
\centering
\includegraphics[width=0.4\textwidth]{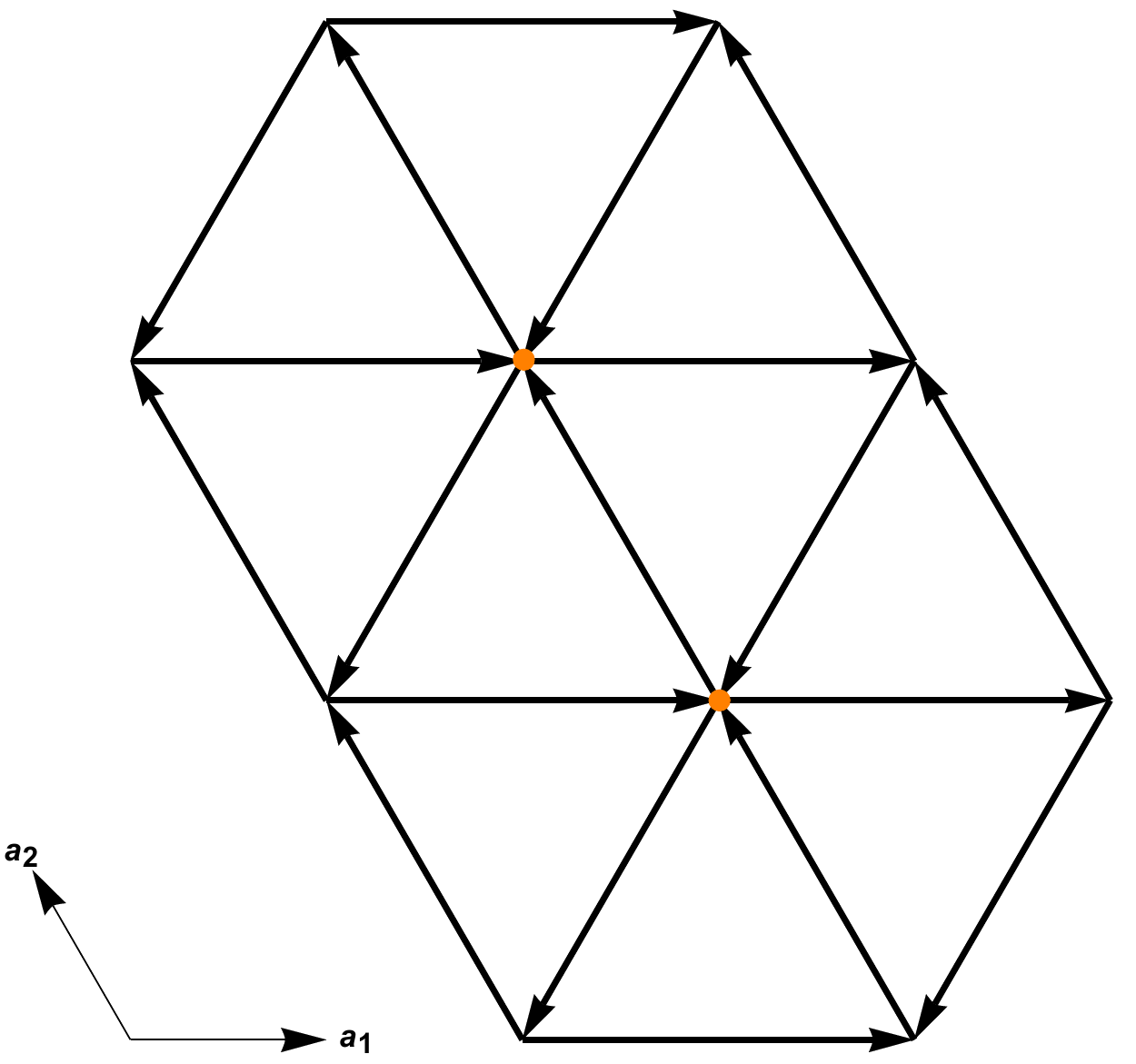}
\caption{Zero flux ansatz for Schwinger fermion with nearest neighbor pairing. The pairing term is odd under inversion. The direction of the arrow denotes the positive pairing. }
\label{fig:fermion_zero_flux}
\end{figure}

 The PSG  for the $f$ particle is $P^f=(0,0,1)$. The hopping term is equal for every bond. The pairing term follows the pattern showed in Fig.~\ref{fig:fermion_zero_flux}.

  The mean field ansatz  is:

\begin{equation}
	H_M=-\mu \sum_{i}f_i^\dagger f_i-t\sum_{\langle ij \rangle}(f_i^\dagger f_j+h.c.)- \sum_{\langle ij \rangle}(\Delta_{ij}^*f_i f_j+h.c.)
\end{equation}

We need to fix the filling $\langle f_i^\dagger f_i \rangle =\frac{1}{6}$. The pairing ansatz is $\Delta_{ij}=-\Delta_{ji}$ with $|\Delta_{ij}|=\Delta$.

\begin{equation}
	H=\frac{1}{2}\sum_{\mathbf k \in BZ}(f^\dagger(\mathbf k), f(-\mathbf k))H(\mathbf k)
	\left(\begin{array}{c}f(\mathbf k)\\f^\dagger(-\mathbf k)\end{array}\right)
\end{equation}

with
\begin{equation}
	H(\mathbf k)=\left(
	\begin{array}{cc}
	\xi(\mathbf k) & \Delta(\mathbf k)^* \\
	\Delta^\dagger(\mathbf k)& -\xi(-\mathbf k)
	\end{array}
	\right)
\end{equation}

Let us define $k_1=\mathbf{k}\cdot \mathbf{a_2}=k_x$, $k_2=\mathbf{k}\cdot \mathbf{a_2}=-\frac{1}{2}k_x+\frac{\sqrt{3}}{2}k_y$ and $k_3=\mathbf{k}\cdot (\mathbf{a_1+a_2})=\frac{1}{2}k_x+\frac{\sqrt{3}}{2}k_y$.

We have 
\begin{equation}
	\epsilon(\mathbf k)=-2t (\cos k_1+\cos k_2+\cos k_3)
\end{equation}

\begin{equation}
	\Delta(\mathbf k)=2 i \Delta(\sin k_1+\sin k_2-\sin k_3)
\end{equation}

The dispersion is
\begin{equation}
	E(\mathbf k)=\sqrt{(\epsilon(\mathbf k)-\mu)^2+4 \Delta^2(\sin k_1+\sin k_2-\sin k_3)^2}
\end{equation}

Self consistently we find $\Delta=3.7522 J$ with mean field energy $E_M=-0.02039 \times 42 J$. $\mu=-7.095 J$.

\end{document}